\documentclass[prb,aps,reprint,twocolumn,groupedaddress,english,eqsecnum,showpacs,longbibliography]{revtex4-1}
\usepackage{babel}
\usepackage{xcolor}
\usepackage{graphicx}
\usepackage{amsmath}
\usepackage{amssymb}

\definecolor{darkblue}{rgb}{0.1,0.2,0.6}
\definecolor{darkred}{rgb}{1,0.5,0}
\usepackage[colorlinks,citecolor=darkblue,linkcolor=darkred,urlcolor=darkblue]{hyperref} 
\usepackage[all]{hypcap} 

\makeatletter
\adddialect\l@English\l@english
\makeatother

\def\be{\begin{equation}} \def\ee{\end{equation}}

\def\bea{\begin{eqnarray}} \def\eea{\end{eqnarray}}

\newcommand{\dagga}{{\phantom{\dagger}}}
\newcommand{\mat}[1]{\boldsymbol{#1}}
\newcommand{\veck}{{\bf k}}

\newcommand{\ie}{\textit{i.e.} }

\newcommand{\vs}{\textit{vs.} }

\newcommand{\tr}{\mathrm{Tr}}

\definecolor{commentcolor}{rgb}{0.1,0.2,0.6}
\definecolor{commentcolor2}{rgb}{1,0,0}
\definecolor{commentcolorF}{rgb}{0,0,1}
\definecolor{commentcolorD}{rgb}{1,0.1,.1}
\definecolor{todocolor}{rgb}{0.4,0,1}

\renewcommand{\vec}[1]{\mathbf{#1}}
\renewcommand{\mat}[1]{\mathbf{#1}}

\begin{document}
\title{Entanglement entropies of the $J_1 - J_2$ Heisenberg antiferromagnet on the square lattice}
\author{Nicolas Laflorencie}
\affiliation{Laboratoire de Physique Th\'eorique, IRSAMC, Universit\'e de Toulouse,
{CNRS, 31062 Toulouse, France}}
\author{David J. Luitz}
\affiliation{Laboratoire de Physique Th\'eorique, IRSAMC, Universit\'e de Toulouse,
{CNRS, 31062 Toulouse, France}}
\author{Fabien Alet}
\affiliation{Laboratoire de Physique Th\'eorique, IRSAMC, Universit\'e de Toulouse,
{CNRS, 31062 Toulouse, France}}

\date{\today}

\begin{abstract}
Using a modified spin-wave theory which artificially restores zero sublattice magnetization on finite lattices, we investigate the entanglement properties of the N\'eel
ordered $J_1 - J_2$ Heisenberg antiferromagnet on the square
lattice. Different kinds of subsystem geometries are studied, either
corner-free (line, strip) or with sharp corners (square). Contributions from
the $n_G=2$ Nambu-Goldstone modes give additive logarithmic corrections with
a prefactor ${n_G}/{2}$ independent of the R\'enyi index. On the other hand,
corners lead to additional (negative) logarithmic corrections with a
prefactor $l^{c}_q$ which does depend on both $n_G$ and the R\'enyi index
$q$, in good agreement with scalar field theory predictions. By varying the second neighbor coupling $J_2$ we also explore universality across the N\'eel ordered side of the phase diagram of the $J_1 - J_2$ antiferromagnet, from the frustrated side $0<J_2/J_1<1/2$ where the area law term is maximal, to the strongly ferromagnetic regime $-J_2/J_1\gg1$ with a purely logarithmic growth $S_q=\frac{n_G}{2}\ln N$, thus recovering the mean-field limit for a subsystem of $N$ sites. Finally, a universal
subleading constant term $\gamma_q^{\rm ord}$ is extracted in the case
of strip subsystems, and a direct relation is found (in the large-S
limit) with the same constant extracted from free lattice systems. The singular limit of vanishing
aspect ratios is also explored, where we identify for $\gamma_q^\text{ord}$ a regular part and a singular component, explaining the discrepancy of the linear scaling term for
fixed width \vs fixed aspect ratio subsystems.

\end{abstract}
\maketitle
%
\section{Introduction}
\begin{figure}[t] \centering \includegraphics[width=\columnwidth,clip]{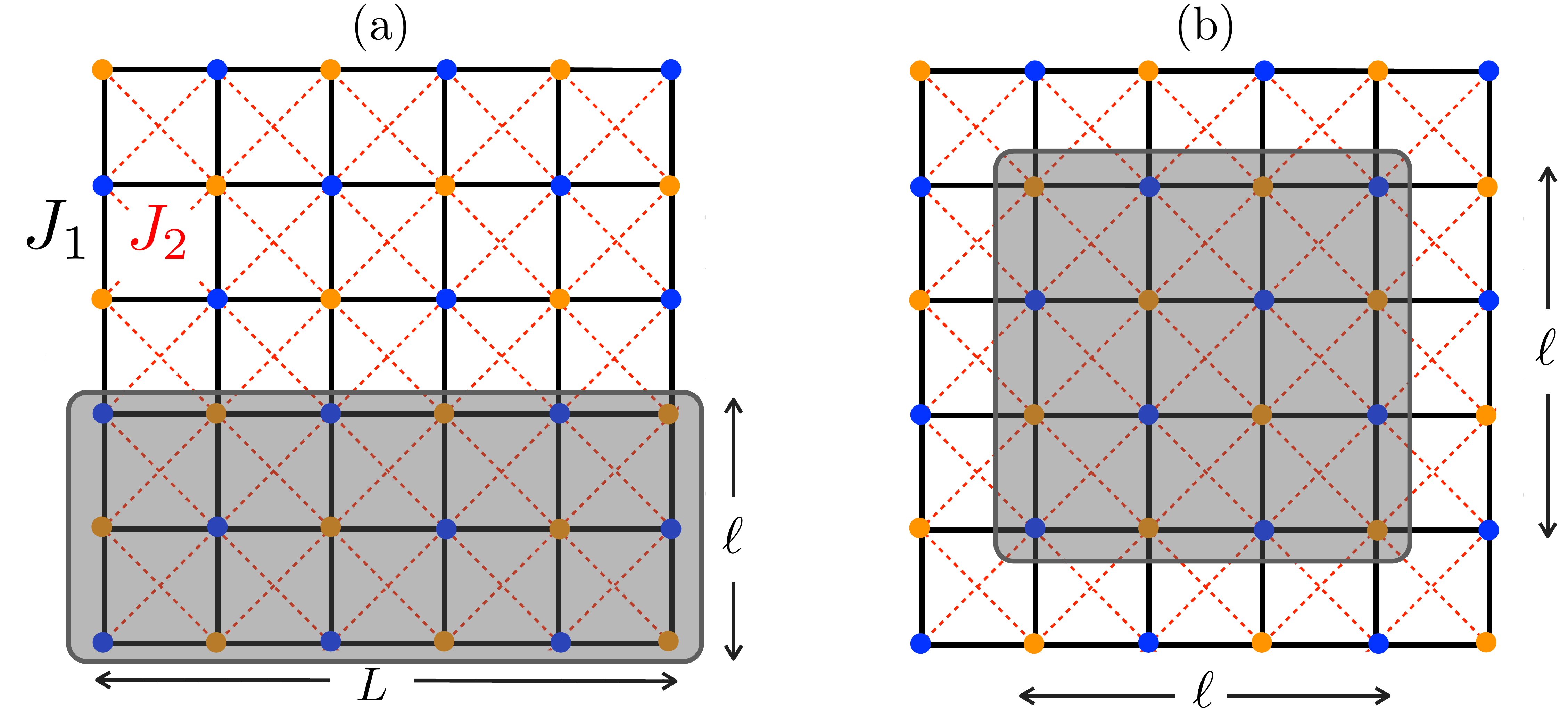}
	\caption{Schematic picture for the $L\times L$ square lattice $J_1 - J_2$ antiferromagnet on a
    torus with 2 types of entanglement bipartitions: (a) rectangular strip of extent $L\times \ell$ with no
corners and (b) square of extent $\ell\times \ell$ with four $\pi/2$
corners. In all this work, we assume periodic boundary conditions in both directions.}
\label{fig:lattice} 
\end{figure}

Entanglement properties of interacting quantum spin systems have recently
attracted a lot of interest. In particular, great attention is paid to the
universal information carried by bipartite entanglement measures such as the R\'enyi entanglement entropies (EEs) defined by
\be
S_q=\frac{1}{1-q}\ln{\rm{Tr}}\left(\hat\rho_{\Omega}\right)^q,
\ee
where $\hat\rho_{\Omega}$ is the reduced density matrix of a given subsystem $\Omega$ (see
Fig.~\ref{fig:lattice}) computed in the ground-state wave-function. Note that the special limit of
$q\to 1$ corresponds to the standard von Neumann EE given by $S_1 = -\tr \rho_\Omega \ln
\rho_\Omega$ and is always implicitly understood whenever we refer to $q=1$.
As a general result, at $T=0$ the R\'enyi EEs follow an area law~\cite{eisert_colloquium:_2010,note_area} in dimension $d$
\be
S_q=a_q L^{d-1} +\cdots
\ee
where $L^{d-1}$ is the size of the boundary between subsystem $\Omega$ and
the rest, and the ellipses are subleading corrections. Such corrections have
been shown to carry universal information about topological
order~\cite{kitaev_topological_2006,levin_detecting_2006,furukawa_topological_2007,isakov_topological_2011},
or the presence of Nambu-Goldstone modes associated to the breaking of a
continuous
symmetry~\cite{metlitski_entanglement_2011,song_entanglement_2011,kulchytskyy_detecting_2015,luitz_universal_2015}. In
the latter case, Metlitski and Grover (MG) \cite{metlitski_entanglement_2011} have derived the following analytical expression in the case of smooth boundaries (no corner), as for instance depicted for $d=2$ in Fig.~\ref{fig:lattice} (a) for $L\times \ell$ strip subsytems:
\be
S_q=a_qL^{d-1}+\frac{n_G}{2}\ln\left(\frac{\rho_{s}}{v} L^{d-1}\right) +\gamma_q^{\rm ord},
\label{eq:1.3}
\ee
where $\rho_s$ is the stiffness, $v$ the velocity of the $n_G$
Nambu-Goldstone modes, and $\gamma_q^{\rm ord}$ a universal geometric
constant. In the case of subsystems having sharp corners, as depicted in
Fig.~\ref{fig:lattice} (b), it is expected that~\cite{metlitski_entanglement_2011}:
\bea
S_q=a_qL^{d-1}&+&\frac{n_G}{2}\ln\left(\frac{\rho_{s}}{v} L^{d-1}\right)\nonumber\\
&+&n_G\sum_c l_q^{c}(\varphi_c)\ln\left(\frac{L}{a}\right)+\gamma_q^{\rm ord},
\label{eq:1.4}
\eea
where $a$ is a non-universal length scale, and the corner contribution depends on $n_G$, the R\'enyi parameter $q$, and the number of corners $c$ of angle $\varphi_c$.
The contributions $l_q^c(\varphi_{c})$ from each corner come from the (free)
Goldstone modes and can be computed, following the work of 
Casini and Huerta~\cite{casini_universal_2007} on scalar field theory, by the numerical solution of a set of
non-linear differential equations, valid for $\varphi_c\in [0,\pi]$
($l_q(\varphi)=l_q(2\pi-\varphi)$) and $q \in \mathbb{N}\setminus \{1\}$. 

Previous works have explored the scaling of the entanglement entropy in
ground-states of systems that break continuous symmetries in the
thermodynamic limit. Subleading logarithmic corrections arising from the
Goldstone modes have been observed in quantum Monte Carlo simulations of finite
spin systems~\cite{kallin_anomalies_2011,humeniuk_quantum_2012,helmes_entanglement_2014}, even though the prefactor of this correction did not
perfectly agree with the prediction $n_G/2$, until a very recent
large-scale, low-temperature quantum Monte Carlo (QMC) investigation by Kulchytskyy {\it et
  al.}~\cite{kulchytskyy_detecting_2015} for the 2d XY
model and $q=2$. Logarithmic corrections have also been observed in
finite-size SW calculations~\cite{song_entanglement_2011} (similar to the ones presented in this
manuscript), but not with a high-enough precision to again ascertain the
prediction, except for the case a line-shaped subsystem for which the
prefactor $n_G/2$ could be recovered assuming further subleading
corrections~\cite{luitz_universal_2015} (see also the recent work Ref.~\onlinecite{frerot_area_2015}). The existence of logarithmic
corrections have also been discussed based on a phenomenological picture of the tower of low-lying
states in the symmetry-broken phase of antiferromagnets~\cite{kallin_anomalies_2011}. Logarithmic
corrections due to corner contributions have on the other hand been
identified and calculated precisely in free lattice systems~\cite{casini_universal_2007},  
broken continuous symmetries systems~\cite{kallin_anomalies_2011} as well as for
various critical points using QMC, cluster expansions or tree tensor network
techniques~\cite{tagliacozzo_simulation_2009,singh_thermodynamic_2012,kallin_anomalies_2011,humeniuk_quantum_2012,kallin_entanglement_2013,inglis_entanglement_2013,kallin_corner_2014,
helmes_entanglement_2014,stoudenmire_corner_2014,devakul_quantum_2014,devakul_entanglement_2014,helmes_correlations_2014}. In a recent work~\cite{bueno_universality_2015},
predictions for the universality of corner contributions in various
theories are also provided. Finally, Kulchytskyy {\it et
  al.}~\cite{kulchytskyy_detecting_2015} could also compute
with QMC the subleading constant correction $\gamma_2^{\rm ord}$ in the 2d
XY model, finding a good agreement with the prediction of MG in
Ref.~\onlinecite{metlitski_entanglement_2011}.

In this paper, we provide a systematic high-precision study of the
universal nature of three subleading terms of the R\'enyi EE  appearing in Eq.~\eqref{eq:1.4} for a generic model of quantum antiferromagnetism in two dimensions ($d=2$). This is achieved using a large-$S$ semi-classical approach, the modified linear spin-wave (SW) theory, where the rotational SU(2) symmetry, while practically broken, is artficially restored for finite size systems~\cite{takahashi_modified_1989,hirsch_spin-wave_1989}. We focus on the $J_1 - J_2$ spin-$S$ antiferromagnet defined on a bipartite $L\times L$ square lattice by the following Hamiltonian
\bea
{\cal H}&=&J_1\sum_{\langle i j\rangle}{\vec{S}}_i\cdot {\vec{S}}_j + J_2\sum_{\langle\langle i j\rangle\rangle}{\vec{S}}_i\cdot {\vec{S}}_j\nonumber\\
&+&h\sum_{i}(-1)^iS_i^z,
\label{eq:J1J2}
\eea
where ${\vec{S}}$ are spin-$S$ operators, interactions act between nearest
neighbours $\langle ij\rangle$ and second nearest neighbours
${\langle\langle i j\rangle\rangle}$ along the diagonals of a square lattice
(see Fig.~\ref{fig:lattice}), and $h$ is an external staggered field. We
impose periodic boundary conditions in all directions.
At $h=0$ this model spontaneously breaks the SU(2) symmetry at zero temperature in the thermodynamic
limit, and displays N\'eel order for $J_2< J_2^c$, with $J_2^c\to J_1/2$ for $S\to
\infty$~\cite{chandra_possible_1988}. The restoration of zero sublattice magnetization {in finite
systems} is made possible by tuning the small staggered field $h^*(L)$ such that on any site $\langle S^z_i\rangle =0$. As first done in Refs.~\onlinecite{song_entanglement_2011,luitz_universal_2015}, this allows to correctly compute R\'enyi EEs on finite systems. Here we make a systematic and extensive study across the full N\'eel regime $-\infty<J_2< J_2^c$ for various subsystem shapes and sizes in order to characterize contributions form (i) Nambu-Goldstone modes, (ii) corners, (iii) frustration effects $J_2/J_1>0$, and (iv) geometric effects appearing through the universal constant $\gamma_q^{\rm ord}$ in Eq.~\eqref{eq:1.3}. 

Let us briefly summarize our main results. Using a large-$S$ approach, we have numerically
extracted the three subleading corrections in the scaling of EEs Eq.~\eqref{eq:1.4} with $n_G=2$ for
SU(2) antiferromagnets. Universality has been tested in the entire N\'eel ordered regime of the $J_1
- J_2$ Heisenberg model Eq.~\eqref{eq:J1J2} for various $S$, even in the
frustrated regime where QMC is inapplicable. In the case of subsystems having sharp corners, small negative
corner terms $l_q^c$ are found, in perfect agreement with the predictions by Casini and Huerta for free
scalar fields~\cite{casini_universal_2007}. The non-universal area-law term has also been studied as a function of the second neighbour
coupling, showing remarkable behaviors both in the mean-field limit ($-J_2/J_1\gg 1$) where it
vanishes, and close to the frustrated critical point $J_2^c$ where the area law prefactor
strongly increases, while log corrections due to Nambu-Goldstone modes are still present. Furthermore, the additional geometric constant
$\gamma_q^{\rm ord}$, which depends on the subsystem aspect ratio $\ell/L$, is extracted for
various R\'enyi indices, and a simple relation with the free scalar field result is derived.  We have also explored the limit of vanishing aspect ratios where a non-trivial slow singular behavior shows up as $\gamma_{q}^{\rm ord}(\ell/L\ll 1)\to -\infty$.

The rest of the paper is organized as follows. In Section~\ref{sec:2} we start by recalling the
modified SW formalism for the $J_1 - J_2$ spin-$S$ antiferromagnet, and how it can be used to
compute the R\'enyi EEs. We then turn to the results for EEs in Section \ref{sec:3} where we discuss
several aspects: we first describe numerical diagonalization results, which can be conveniently
performed up to subsystems of $\lesssim 10^5$ sites, for various shapes of
subsystems including strips (Sec.~\ref{sec:strip} and
Fig.~\ref{fig:lattice}a) and squares (Fig.~\ref{fig:lattice}b), with a particular
focus on the corner contributions (Sec.~\ref{sec:corners}) and their dependence on
the R\'enyi parameter $q$. In Section~\ref{sec:area} the dependence on the second neighbour coupling
$J_2$ is studied, focussing on the non-universal area law prefactor $a_q$.
In Section \ref{sec:4} we discuss the constant term $\gamma_q^{\rm ord}$
which is compared to the field-theory prediction of
MG in Sec.~\ref{sec:gamma1}. An interesting connection to the free scalar field result is achieved in Sec~\ref{sec:gamma2}. We further explore the singular limit of vanishing aspect ratios in Sec.~\ref{sec:gamma3} using quasi-analytical results
for single and double line subsystems where translation symmetry inside the subsystem
allows to get an explicit expression for $S_q$. 
Finally we summarize and discuss our results in Section \ref{sec:5}. Details of spin-wave calculations are provided in
Appendix~\ref{sec:appA}, analytical results for the mean-field limit
$-J_2/J_2\gg 1$ are presented in Appendix~\ref{sec:MF}, and an analytical derivation for one-dimensional subsystems is given in Appendix~\ref{sec:1d}.

\section{Modified spin-wave approach}
\label{sec:2}

\subsection{Dyson-Maleev transformation and Bogoliubov diagonalization}

We use the Dyson-Maleev formalism~\cite{dyson_general_1956,maleev__1958} to
map spin operators onto bosonic ones. For sites on sublattice A of the
square lattice:
\be
S_i^z=S-b^{\dagger}_{i}b^\dagga_i,~
S_i^+=(2S-b^{\dagger}_{i}b^\dagga_i)b^\dagga_i,~
S_i^-=b^\dagger_i,
\label{eq:DM1}
\ee
and for the sublattice B:
\be
S_i^z=b^{\dagger}_{i}b^\dagga_i-S,~
S_i^+=-b^\dagger_i(2S-b^{\dagger}_{i}b^\dagga_i),~
S_i^-=-b^\dagga_i.
\label{eq:DM2}
\ee
Truncating at $1/S$ order, the $J_1 - J_2$ Hamiltonian Eq.~\eqref{eq:J1J2}
becomes (up to a constant)
\bea
{\cal H}&=&\sum_i\left(h+4S[J_1-J_2]\right)b^{\dagger}_{i}b^\dagga_i\\
\label{eq:quad}
&+&\sum_{\langle ij\rangle}SJ_2\left(b^{\dagger}_{i}b^\dagga_j+b^{\dagger}_{j}b^\dagga_i\right)-\sum_{\langle ij\rangle}SJ_1\left(b^{\dagga}_{i}b^\dagga_j+b^{\dagger}_{j}b^\dagger_i\right).\nonumber
\eea
After a Fourier transformation, it reads
\be
{\cal H}=\sum_{\bf k}A_{\bf k}(b^{\dagger}_{\bf k}b^{\dagga}_{\bf k}+b^{\dagger}_{-\bf k}b^{\dagga}_{-\bf k})
+B_{\bf k}(b^{\dagger}_{\bf k}b^{\dagger}_{-\bf k}+b^{\dagga}_{\bf k}b^{\dagga}_{-\bf k}),
\label{eq:Hk}
\ee
with 
\bea
A_{\bf k}&=&2SJ_2\cos k_x\cos k_y+2S(J_1-J_2)+\frac{h}{2}\\
B_{\bf k}&=&-SJ_1\left[\cos k_x + \cos k_y\right].
\label{eq:AkBk}
\eea
The quadratic part of the above Hamiltonian can be diagonalized via a standard Bogoliubov transformation:
\begin{equation}
\begin{array}{ccc}
 b_\veck=u_\veck\alpha_\veck-v_\veck\alpha_{-\veck}^\dagger & & b_\veck^\dagger=u_\veck\alpha_\veck^\dagger-v_\veck\alpha_{-\veck}.
\end{array}
\end{equation}
The quasiparticle operators $\alpha_\veck^\dagga$ and $\alpha_\veck^\dagger$ satisfy bosonic commutation relations provided $u_\veck^2-v_\veck^2=1$, and diagonalize \eqref{eq:Hk} if
\bea
u_\veck^2&=&\frac{1}{2}\left(\frac{A_\veck}{\sqrt{A_\veck^2-B_\veck^2}}+1\right)\\
v_\veck^2&=&\frac{1}{2}\left(\frac{A_\veck}{\sqrt{A_\veck^2-B_\veck^2}}-1\right).
\eea
In terms of Bogoliubov quasi-particles, the $J_1 - J_2$ Hamiltonian takes the simpler form
\be
{\cal H}=\sum_{\bf k}\Omega_{\bf k}\alpha_{\bf k}^{\dagger}\alpha_{\bf
  k}^{\dagga} + {\rm constant},
\label{eq:Hkdiag}
\ee
with the SW excitation spectrum
$\Omega_{\bf k}=2\sqrt{A_{\bf k}^2 -B_{\bf k}^2}$ (this spectrum is illustrated in
Appendix~\ref{sec:appA}). In the vicinity of the two minima at ${\bf k}=(\pi,\pi)$ and $(0,0)$, the dispersion is linear, with a velocity
\be
v_{\rm sw}=2\sqrt{2}S\sqrt{J_1(J_1-2J_2)},
\label{eq:vsw}
\ee
which is defined only on the AF side $J_2<J_1/2$. The SW spectrum and
velocity are illustrated in Fig.~\ref{fig:SW} and ~\ref{fig:map} of Appendix~\ref{sec:appA}.

In the thermodynamic limit, the continuous SU(2) symmetry of the original $J_1 -
J_2$ Hamiltonian can be spontaneously broken, with the two associated
Nambu-Goldstone modes at ${\bf k}=(\pi,\pi)$ and $(0,0)$. The
corresponding staggered magnetization order parameter is given at the $1/S$ order by
\bea
m_{\rm AF}&=&\lim_{h\to 0}\lim_{N\to\infty}|\langle S_i^z\rangle|\nonumber\\
&=&S+\frac{1}{2}-\frac{1}{8\pi^2}\int_{\rm Bz}{\rm{d}}^2{\bf k}\frac{A_{\bf k}}{\Omega_{\bf k}}.
\label{eq:mafsw}
\eea
In Appendix~\ref{sec:appA}, this expression is evaluated numerically to
obtain the range of parameter space where N\'eel order is expected from this
SW treatment.

\subsection{Spin-Wave theory for finite size systems}
The above SW approach assumes a classical ordered state as a starting
point. This does not allow for a correct study of finite size effects since
the spin rotational symmetry has to remain unbroken on finite-size
lattices. In order to repair this, adding a staggered magnetic field to the
quantum antiferromagnet allows to artificially restore zero sub-lattice
(SW-corrected) magnetization, as originally proposed in
Refs.~\onlinecite{takahashi_modified_1989,hirsch_spin-wave_1989}. This will
turn crucial to capture the subleading scaling terms in the entanglement entropy. 

\begin{figure}[t] \centering \includegraphics[width=\columnwidth,clip]{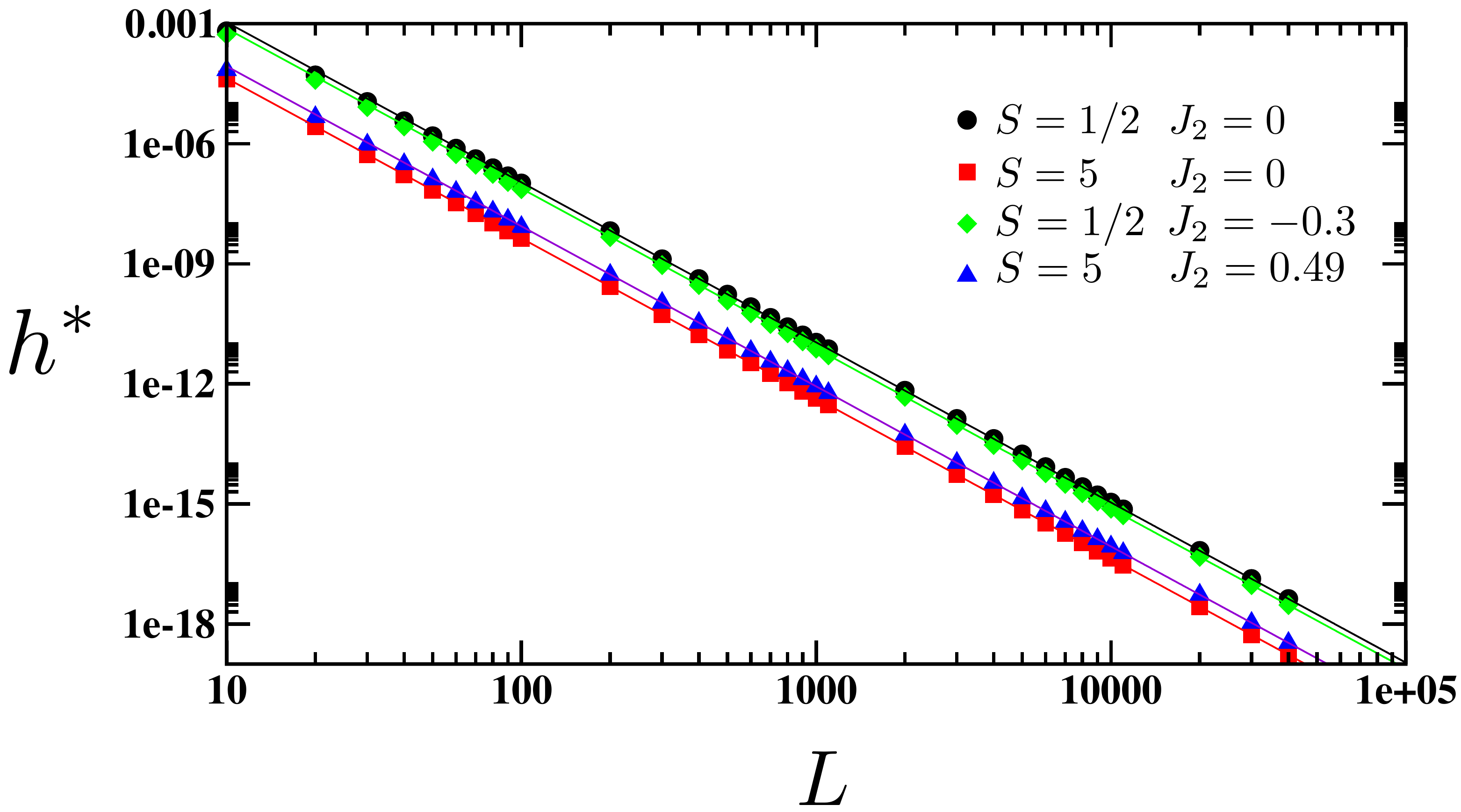}
	\caption{Regularizing staggered magnetic field $h^*$ in the $J_1 - J_2$ antiferromagnet such that SW corrected magnetization vanishes.}
\label{fig:hstar} \end{figure}

In this approach, one imposes that for any given finite size sample $\langle S_i^z\rangle=0\quad \forall i$,
which yields a staggered field $h^*$ such that the number of Holstein-Primakoff bosons $\langle n\rangle = S$. This leads to
\be
\sum_{\bf k}\frac{A_{\bf k}(h^*)}{\Omega_{\bf k}(h^*)}=N(2S+1).
\label{eq:sum}
\ee
This regularizing field is very small and scales rapidly to zero with the system size~\cite{song_entanglement_2011}. Indeed, one can rewrite Eq.~\eqref{eq:sum} as follows
\be
N(2S+1)-\sum_{{\bf k}\neq {\bf k}_0}\frac{A_{\bf k}(h^*)}{\Omega_{\bf k}(h^*)}
=2\frac{A_{{\bf k}_0}(h^*)}{\Omega_{{\bf k}_0}(h^*)},\ee
where ${\bf k}_0=(0,0)$ and $(\pi,\pi)$ are the singular modes where the dispersion vanishes in the absence of staggered field. The contributions from these two modes, divergent in the limit $h^*\to 0$, are similar:
\be
\frac{A_{{\bf k}_0}}{\Omega_{{\bf k}_0}}=\frac{4SJ_1+h^*}{\sqrt{h^*(h^*+8SJ_1)}}.
\ee
Defining 
\be
m^*(N,h^*)=S+\frac{1}{2}-\frac{1}{2N}\sum_{{\bf k}\neq {\bf k}_0}\frac{A_{\bf k}(h^*)}{\Omega_{\bf k}(h^*)}
\ee
we obtain a self-consistent equation for $h^*$
\be
h^*=4SJ_1\left[\frac{1}{\sqrt{1-\left(\frac{1}{Nm^*}\right)^2}}-1\right].
\label{eq:sch}
\ee
In the limit $N\gg 1$, $m^*\to m_{\rm AF}$, and we have
\be
h^*\simeq \frac{2SJ_1}{m_{\rm AF}^2}\frac{1}{N^2}.
\label{eq:hstar}
\ee
As seen below, it is essential to determine the actual value of $h^*(L)$ with a
high precision  in order to compute accurately various finite size
correlations. Since the field $h^*$ gets rapidly very small with increasing
system sizes, we resort to a multiple precision 
evaluation of the self-consistent equation Eq.~\eqref{eq:sch}. In Fig.~\ref{fig:hstar} we present the result showing the behavior of $h^*(L)$ for some representative values of $J_2$ and $S$. In all cases, the staggered field vanishes very fast and is well described by Eq.~\eqref{eq:hstar} at large enough $L$.

Interestingly this small staggered field opens a gap in the excitation spectrum 
\bea
\Delta^*&\simeq& \sqrt{2SJ_1 h^*}\nonumber\\
&\simeq& \frac{2SJ_1}{m_{\rm AF}}\frac{1}{N}
\eea
which scales in the same way as the Anderson tower of 
states~\cite{anderson_approximate_1952}. 
Therefore, the excitation spectrum has linearly dispersing Nambu-Goldstone (SW) modes with a level
spacing $\sim 1/L$ and a tower of states like finite size gap $\sim 1/L^2$ produced by the symmetry restoring staggered field. 

We use this modified finite-size SW approach to compute the entanglement
entropy as detailed below. In order to show that it reproduces fairly well
the physics of finite-size systems, we also compare in
Appendix~\ref{sec:appA} results for the finite-size structure factor for
$S=1/2$ and various $J_2<0$ to the ones obtained with the exact QMC method.

\subsection{Entanglement entropy}

As the diagonalized Hamiltonian Eq.~\eqref{eq:Hkdiag} is non-interacting, Wick's
theorem eases the computation of entanglement entropy, which can nicely be
extracted from the correlation
matrix~\cite{peschel_reduced_2009}, an object which
contains all two-body correlations within a block of sites. For completeness, we
recapitulate here the essential formulae.

We first need to define single particle Green's function $\langle b^{\dagger}_{i}b^\dagga_j\rangle=-\frac{\delta_{ij}}{2}+f_{ij}$ and $\langle b^{\dagga}_{i}b^\dagga_j\rangle=g_{ij}$, with
\bea
f_{ij} & = & \frac{1}{2N}\sum_{\bf k}\frac{A_{\bf k}\cos\left[\bf
    k\cdot({\bf r}_{i}-{\bf r}_j)\right]}{\Omega_{\bf k}}\nonumber \\
g_{ij} & = & -\frac{1}{2N}\sum_{\bf k}\frac{B_{\bf k}\cos\left[\bf k\cdot({\bf r}_{i}-{\bf r}_j)\right]}{\Omega_{\bf k}}.
\label{eq:fijgij}
\eea
We remark that $g_{ij}=0$ ($f_{ij}=0$) if $i$ and $j$ belong to the same
(different) sublattice(s).

The entanglement entropy of a region $\Omega$ containing $N_\Omega$ sites
can then be
extracted~\cite{bombelli_quantum_1986,plenio_entropy_2005,barthel_entanglement_2006}
from the eigenvalues $\nu_l^2$ of the $N_\Omega\times N_\Omega$ correlation matrix ${\bf{C}}$
\be
	C_{ij} = \sum_{i'\in \Omega} (f_{ii'}+g_{ii'})(f_{i'j}-g_{i'j})
 \label{eq:Crr}
\ee 
where $i,j\in \Omega$. Due to the sublattice properties of $f$ and $g$, we
have that $C_{ij}=C_{ji}$ if $i$ and $j$ belong to the same sublattice,
$C_{ij}=-C_{ji}$ otherwise. 

The R\'enyi entanglement entropy is obtained as~\cite{peschel_reduced_2009}
\be
	{S}_q =\frac{1}{q-1} \sum_{l=1}^{N_\Omega} \ln\left[\left(\nu_l+\frac{1}{2}\right)^q-
	\left(\nu_l-\frac{1}{2}\right)^q\right],
\ee
which for $q=1$ reads
\be
S_1=\sum_{l=1}^{N_\Omega}\sum_{\epsilon=\pm 1}
\epsilon\left(\nu_l+\frac{\epsilon}{2}\right)\ln\left(\nu_l+\frac{\epsilon}{2}\right),
\ee
and for $q=\infty$
\be
S_{\infty}=\sum_{l=1}^{N_\Omega} \ln\left(\nu_l+\frac{1}{2}\right).
\ee
As first shown by Srednicki~\cite{srednicki_entropy_1993}, Callan and
Wilczek~\cite{callan_geometric_1994}, the entropy of a free massless bosonic
field obeys a strict area law, which is what we observe (data not shown) in the absence of the regularizing staggered field $h^*$. However, as we will see below, the finite staggered field which opens a finite size gap $\sim 1/N$ leads to an additive logarithmic correction proportional to the number of Goldstone bosons.

\section{Results for EE}
\label{sec:3}
\subsection{Strip geometry}
\label{sec:strip}
\begin{figure}[t]
    \centering
    \includegraphics{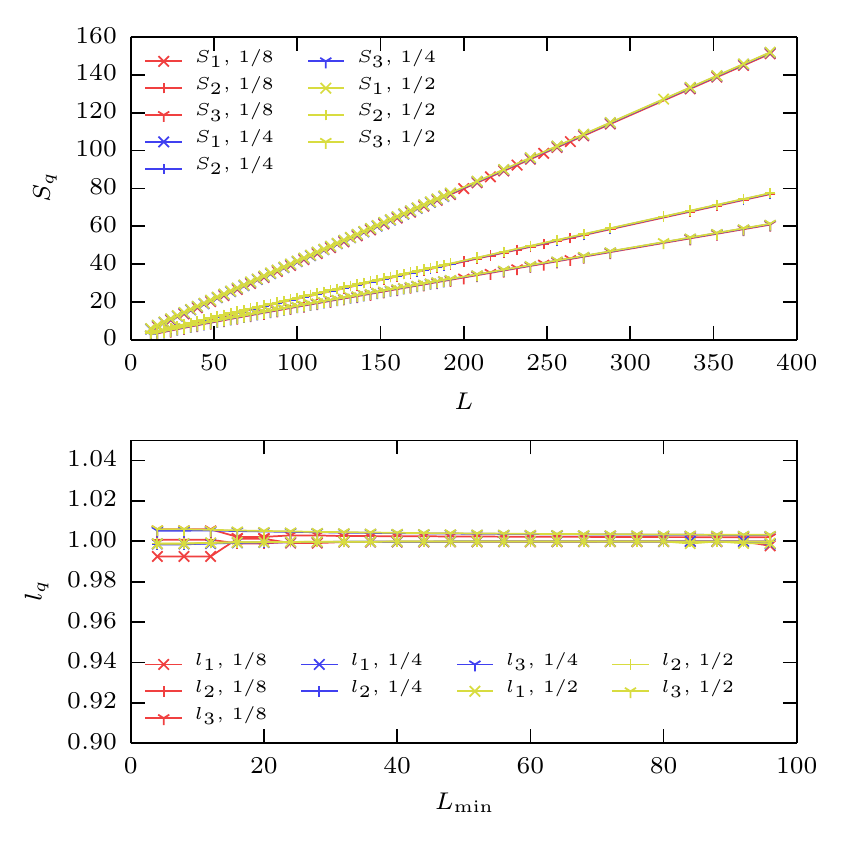}
    \caption{Entanglement R\'enyi entropies for the strip subsystem with different aspect ratios
        $\ell/L$ (upper panel) and fit results for the prefactor $l_q$ of the logarithmic scaling
        term as a function of the minimal system size $L_\text{min}$ included in the fit (lower
    panel). The results displayed here have been obtained for $S=\frac 1 2$ and $J_2=0$. Clearly, $l_q=1$ independent of $q$ and the aspect ratio of the subsystem.}
    \label{fig:strip_raw}
\end{figure}

Let us start with the case of an $L\times \ell$ strip subsystem embedded in an $L\times L$ torus, as
depricted in Fig.~\ref{fig:lattice} (a). This geometry has no corner and we therefore expect the
expression Eq.~\eqref{eq:1.3} to hold. Results obtained from the exact diagonalization of the
correlation matrix ${\bf C}$ for systems up to $\sim10^5$ lattice sites are shown in the upper
panel of Fig.~\ref{fig:strip_raw} where the R\'enyi entropies for $q=1,2,3$ are displayed
for three representative aspect ratios $\ell/L=1/2,~1/4,~1/8$.  Note that
for this strip geometry, translation symmetry of the subsystem is used,
allowing the diagonalization procedure to reach large sizes. This plot clearly demonstrates the
area law behavior $S_q\sim a_q L$ since the dominant scaling behavior does not depend on the number of
subsystem sites but only on its perimeter $2 L$, which is independent of the aspect ratio of the
subsystem. The properties of the area law prefactor $a_q$
will be analyzed in detail in Sec.~\ref{sec:area}, and the universal additive constant $\gamma_q$
from Eq.~\eqref{eq:1.3} in Sec.~\ref{sec:gamma}.

Here, we want to focus on the logarithmic correction associated to the breaking of
SU(2) rotational symmetry with $n_{G}=2$ Nambu-Goldstone modes, expected to be $\frac{n_{G}}{2}\ln L$. This correction is believed to be
universal as it should not depend on the geometry and only reflect the nature of the 
continuous symmetry which is broken in the ground state
\cite{metlitski_entanglement_2011,kulchytskyy_detecting_2015,luitz_universal_2015}. Therefore, we perform fits to the general scaling ansatz
\begin{equation}
    S_q(L) = a_q L + l_q \ln L + b_q + c_q/L +d_q/ L^2,
    \label{eq:EEscaling}
\end{equation}
over various fit ranges $[L_\text{min},L_\text{max}]$. 
Results for $l_q$ are plotted in the lower panel of Fig.~\ref{fig:strip_raw} for various values of the R\'enyi parameter and several aspect ratios.
For $q=1,2$ we clearly observe that $l_q=1$ over basically the whole range of
$L_\text{min}$, whereas for larger values of $q$, the convergence is relatively slow as these
results are to our experience hampered by more severe finite size effects. Nevertheless, the
resulting $l_q$ is already very close to $1$ and the deviation decreases slowly as $L_\text{min}$ is
increased. This leads us to the conclusion that, within our SW approach, we find $l_q=n_G/2=1$ to
be independent on $q$ and the aspect ratio of the subsystem, in perfect
agreement with the field theoretical result by MG~\cite{metlitski_entanglement_2011}.

\subsection{Square subsystems: corner contributions}
\label{sec:corners}
\begin{figure}[b]
    \centering
    \includegraphics{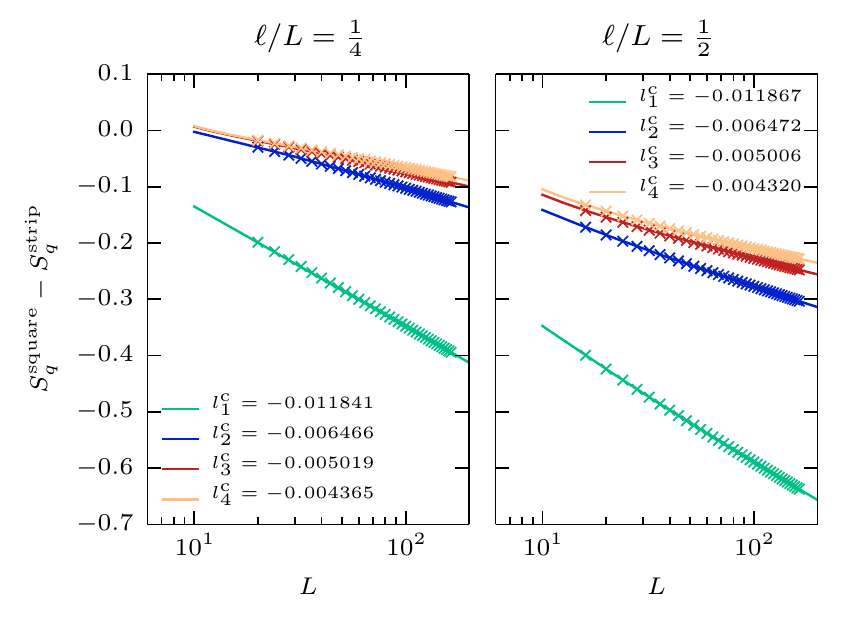}
    \caption{Difference of entanglement entropies for the $S=1/2$ $J_2=0$ Heisenberg antiferromagnet of square and strip subsystems having the same
    boundary length. The remaining dominant scaling term is the logarithmic term which stems from
the corners of the square subsystem. We show fits (full lines) to the form $8l_q^c(\pi/2)\ln(L)+b_q + c_q/L +
d_q /L^2$. SW results (symbols) are shown for two different aspect ratios a for $q=1, 2, 3$ and $4$. }
    \label{fig:corner}
\end{figure}

In addition to the breaking of continuous symmetries, logarithmic corrections to the area law can
also be caused by geometry: In particular, 
logarithmic corrections induced by sharp corners of the subsystem have been discussed in several works~\cite{fradkin_entanglement_2006,casini_universal_2007,
tagliacozzo_simulation_2009,swingle_mutual_2010,singh_thermodynamic_2012,humeniuk_quantum_2012, kallin_entanglement_2013,kallin_corner_2014,
helmes_entanglement_2014,stoudenmire_corner_2014,devakul_quantum_2014,bueno_universality_2015}. The
prefactor of the logarithmic corner correction term is expected to be universal for
all systems with the same type of symmetry breaking/phase transition. However, such corrections are quite difficult to capture with QMC since the prefactor is very small. Together with the contributions coming from
Nambu-Goldstone modes Eq.~\eqref{eq:1.4}, we expect a total correction of the form
\be
{n_G}\left(\sum_c l^c_q(\varphi_c)+\frac{1}{2}\right)\ln L,
\ee
where the sum is taken over all sharp corners inside the subsystem making an
angle $\varphi_c$. Here we aim at numerically extracting $l^c_q(\pi/2)$ for a square subsystem (panel (b) of
Fig.~\ref{fig:lattice}),
expected to coincide with the result of a free scalar
field~\cite{casini_universal_2007}. To do so, we work with a
$L\times L$ torus and substract the entropies of a periodic (corner-free)
strip of size $L\times\ell$ from those of a $L/2\times L/2$ square. Both subsystems
having the same area law $\sim 2L$, independent of the strip aspect ratio
$\ell/L$, and identical logarithmic corrections due to Goldstone modes, we therefore
expect the leading term of this difference to be given only by the corner log contribution:
\be
S^{\rm square}_q-S^{\rm strip}_q=8l_q^c({\pi}/{2})\ln L +\cdots\label{eq:diff}\ee
\begin{figure}[b]
    \centering
    \includegraphics{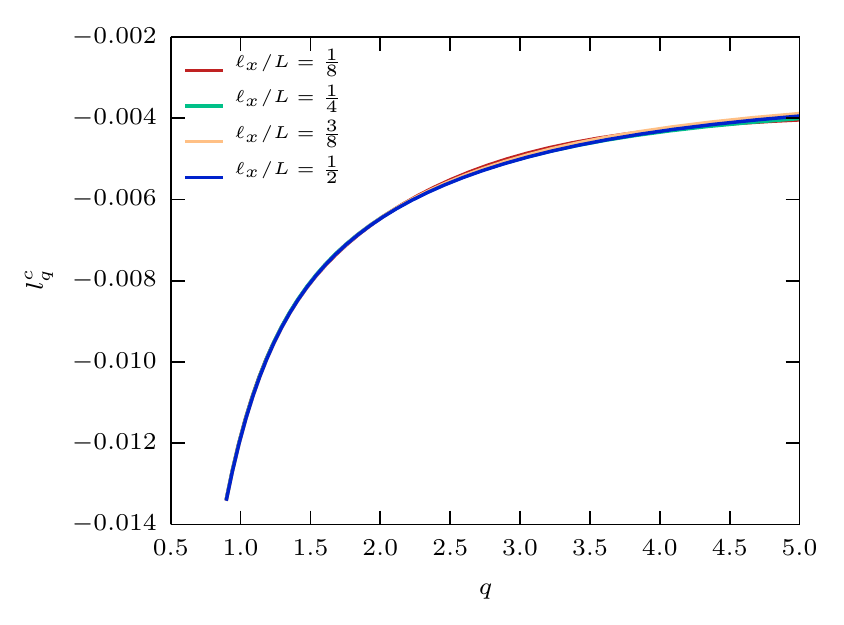}
    \caption{Logarithmic contribution of a $\frac \pi 2$ corner of the subsystem as a function of
    $q$. This result is obtained for $J_2=0$ and $S=\frac 1 2$ by subtracting the entanglement
entropy of a strip subsystem with the same perimeter as the square subsystem with $4$ $\frac \pi 2$
corners and fitting to the same form as shown in Fig. \ref{fig:corner} for different aspect ratios of
the strip. Up to slight deviations for larger R\'enyi indices $q$ due to finite size effects, the
results do not depend on the aspect ratio.}
    \label{fig:cornerlog}
\end{figure}
%
\begin{table}[t]
\begin{tabular}{l|l|l|l|l}
&$q=1$&$q=2$&$q=3$&$q=4$\\
\hline
CH~\cite{casini_universal_2007}&$-0.0118$&$-0.0064$&$-0.0051$&\\
This work&$-0.0118(1)$&$-0.0064(1)$&$-0.0050(1)$&$-0.0043(1)$\\
\hline
\hline
\end{tabular}
\caption{\label{tab:2} Prefactor $l_q^c(\pi/2)$ of the corner logarithmic correction obtained after fitting data in Fig.~\ref{fig:corner}. A comparison with data of Casini and Huerta (CH)~\cite{casini_universal_2007} is also given.}
\end{table}
%
\noindent Numerical results are plotted in Fig.~\ref{fig:corner} where we clearly see
that the above difference Eq.~\eqref{eq:diff} is indeed dominated by a
logarithmic scaling which allows us to extract $l_q^c(\pi/2)$. Small variations of the
results for different aspects ratios of the strips (see left and right
panels of Fig.~\ref{fig:corner}) can be used as a measure of the error due to finite
size effects and fitting procedure. 
Our results are displayed in
Table~\ref{tab:2} where we compare to the free-field results by Casini and Huerta (CH)~\cite{casini_universal_2007}.

Interestingly, we can also study the dependence on the R\'enyi index for
non-integer values of $q$. In Fig.~\ref{fig:cornerlog} we show
$l_q^c(\pi/2)$ versus the R\'enyi parameter $q$ for four different aspect
ratios. For $q$ not too large, the estimates obtained after fitting our numerical data (see caption of Fig.~\ref{fig:corner}), are clearly independent of the aspect ratio, as expected.  This non-trivial $q$-dependence for a free scalar field can be compared to recent numerical results for O(1) and O(2) Wilson-Fisher critical points~\cite{stoudenmire_corner_2014}, featuring qualitatively similar behaviors.
\subsection{$J_2$-dependence and area law prefactor}
\label{sec:area}
\begin{figure}[b]
    \centering
    \includegraphics{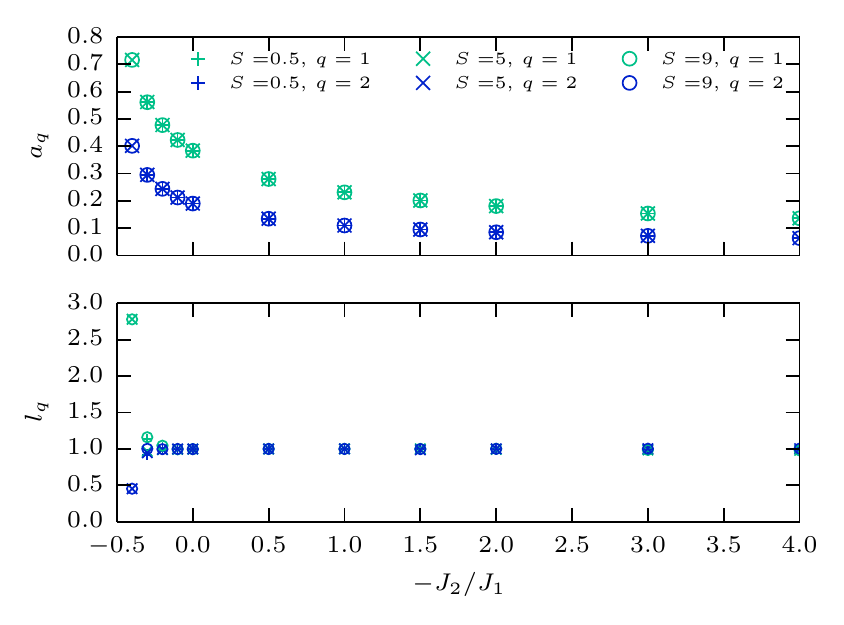}
    \caption{Top: Area law coefficients $a_q$ for $q=1,2$ (extracted for an
      aspect ratio $\ell/L=1/8$) as a function of the next neighbor coupling $J_2/J_1$.
    Approaching the critical point in the frustrated regime ($J_2>0$) the area law coefficient
grows rapidly. Bottom: Prefactor $l_q$ of the log correction in Eq.~\eqref{eq:1.3} due to the
two Goldstone modes of the antiferromagnet. The deviation of $l_q$ from 1 close to the
critical point in the frustrated regime reveals the limitation of the SW
approximation.}
    \label{fig:arealaw}
\end{figure}

Besides universal contributions arising from Nambu-Goldstone modes and
corners, we now study the dominant part which governs the entanglement
growth with the subsystem area. As already discussed in the beginning of the
paper, the $J_1 - J_2$ spin-$S$ Heisenberg model on the square lattice is
N\'eel ordered for $J_2/J_1<0.5$ in the large $S$ limit (see
Appendix~\ref{sec:appA} and Fig.~\ref{fig:mafsw} for the critical value of
$J_2$ as a function of $S$). Scanning across the entire N\'eel ordered regime, we have performed fits to the form Eq.~\eqref{eq:EEscaling} for various values of the second neighbor coupling $J_2$ and spin $S$ for the strip geometry (corner-free) with a $1/8$ aspect ratio. 
Shown in Fig.~\ref{fig:arealaw}, the area law coefficient $a_q$ displays a quite remarkable behavior. First, the results appear to be almost independent of the spin size $S$. Then, as expected from the mean-field limit $J_2/J_2\to -\infty$ 
(see Refs.~\onlinecite{vidal_entanglement_2007,ding_block_2008} and
Appendix~\ref{sec:MF}), $a_q$ goes to zero in the limit $-J_2/J_1\gg
1$. This is because the ground-state becomes more and more classical, with a
very low entanglement. However, less is known when frustration is turned on,
and we observe a rapid growth of the area law term when the critical point
is approached, a feature also observed for the unfrustrated Heisenberg bilayer~\cite{helmes_entanglement_2014}. Note that the validity of the SW calculation can be
questioned when quantum fluctuations become large, approaching the critical
point $J_2^c$.  Nevertheless, we believe the results to be under control if
$1-m_{\rm AF}/S\ll 1$, a condition which can be checked in
Fig.~\ref{fig:mafsw} of Appendix~\ref{sec:appA}. Moreover, as long as the
logarithmic term in the entanglement entropies scaling is still present and
equal to one (due to the two Nambu-Goldstone modes, as shown in the bottom panel of
Fig.~\ref{fig:arealaw} thus confirming the universality across the ordered
phase), we believe the SW approximation correctly captures the behavior of
EE.  In practice, we start to see a deviation of $l_q$ from unity only for the very last points at $J_2/J_1\ge 0.3$ in Fig.~\ref{fig:arealaw}.

\section{Universal additive constant $\gamma^{\rm ord}_q$}
\label{sec:4}
\label{sec:gamma}
\subsection{Direct extraction from large-$S$ data}
\label{sec:gamma1}
\begin{figure}[b]
    \centering
    \includegraphics{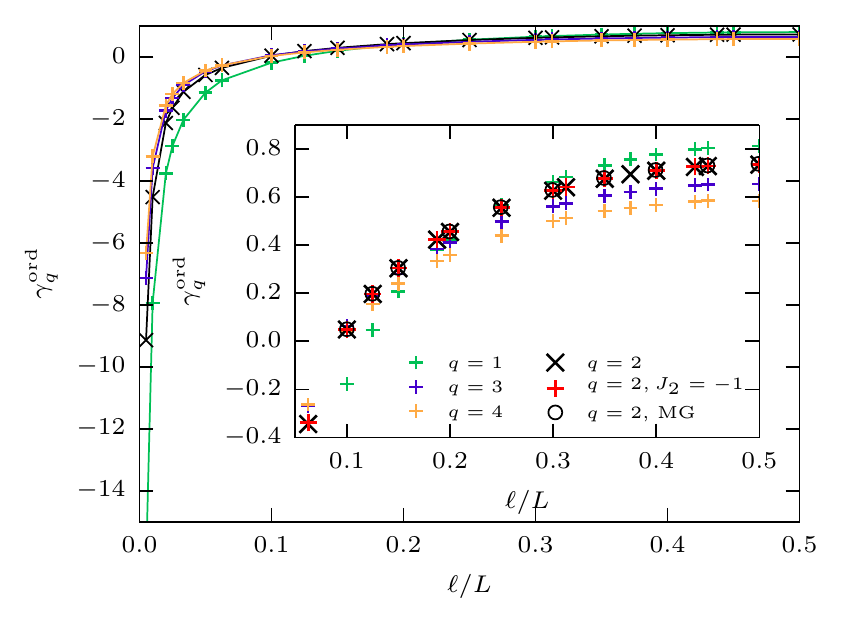}
    \caption{Universal additive constant $\gamma_q^{\rm ord}(\ell/L)$ for different R\'enyi indices $q$ as a
    function of the aspect ratio of the strip geometry for the $S=100$
    Heisenberg $J_1 - J_2$ antiferromagnet at $J_2=0$. The inset displays a
    zoom, showing that our result is in perfect agreement with the universal geometric constant
    obtained by MG~\cite{metlitski_entanglement_2011} for $q=2$.  Results for
    $J_2/J_1=-1$ at $q=2$, shown by red plus signs, agree perfectly with MG (black circles) and
$J_2=0$ (black x), confirming universality.}
    \label{fig:gamma}
\end{figure}

Following MG~\cite{metlitski_entanglement_2011}, a universal additive
constant $\gamma_q^{\rm ord}(\ell/L)$, depending on the aspect ratio
$\ell/L$ of a strip, appears in the R\'enyi EE scaling
Eq.~\eqref{eq:1.3}. Nevertheless, there is also a non-universal term
involving the spin stiffness and the SW velocity in Eq.~\eqref{eq:1.3}. It
is therefore much easier to work in the $S\to\infty$ limit of the $J_1
- J_2$ Heisenberg Hamiltonian where $\rho_s$ and $v$ are known exactly. In
such a limit and having shown above that the logarithmic prefactor is exactly given by $l_q=n_G/2=1$
(the corner constribution vanishes for this geometry), we expect the EE scaling for strips with an aspect ratio $\ell/L$ to be in the $S\gg 1$ limit
\be
S_q=a_qL+\ln\left(S\sqrt{\frac{J_1-2J_2}{8J_1}}\,L\right)+\gamma_q^{\rm ord}(\ell/L).
\label{eq:gammaord}
\ee
This
large $S$ expression has been used to fit our numerical SW data
obtained for $S=100$ and $J_2/J_1=0, -1$. Results for the additive constant
$\gamma_q^{\rm ord}(\ell/L)$ are plotted in Fig.~\ref{fig:gamma} as a
function of the aspect ratio $\ell/L$ for various R\'enyi parameters
$q$. The agreement with the result extracted from
Ref.~\onlinecite{metlitski_entanglement_2011} is excellent for $q=2$. The
universal character of $\gamma_q^{\rm ord}(\ell/L)$ is also corroborated by
the fact that our estimates do not depend on the values of the second
neighbour coupling $J_2/J_1=0,-1$; the sole $J_2$ dependence being given by
the two first terms in Eq.~\eqref{eq:gammaord}.
\begin{figure}[b]
    \centering
    \includegraphics{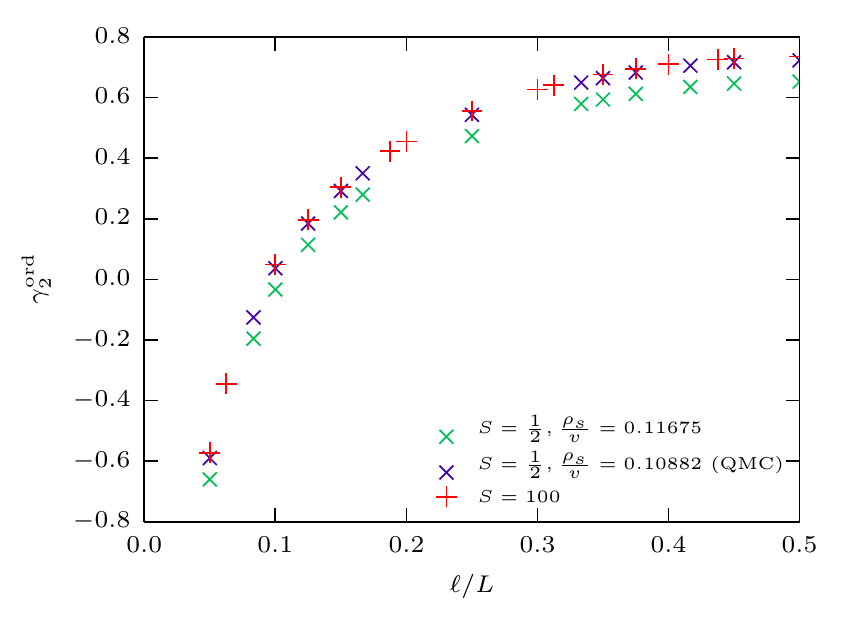}
    \caption{Results for $S=\frac{1}{2}$ and $J_2=0$ for $\gamma_2^\text{ord}$ obtained from
    fits using rectangular subsystems with fixed aspect ratios. For $S=\frac{1}{2}$, we show two
    sets of results using slightly different estimates for $\rho_s/v$. Data shown in blue use
    $\rho_s/v=0.10882(4)$ from the most recent QMC estimate~\cite{sen_velocity_2015}, while data
    shown in green use $\rho_s/v=0.11675$ from a $1/S$ calculation~\cite{hamer_spin-wave_1994}.  The use of the
    QMC result leads to a much better agreement with our $S=100$ data.}
    \label{fig:gamma_comp}
\end{figure}

For the $S=1/2$ Heisenberg model ($J_2=0$), while we have seen above that the logarithmic
corrections are perfectly captured, a precise determination of the additional constant
$\gamma_q^{\rm ord}(\ell/L)$ is less obvious. Indeed, as shown in Fig.~\ref{fig:gamma_comp} for
$q=2$, using $\rho_s/v=0.11675$ from previous $1/S$ estimates~\cite{hamer_spin-wave_1994}, the
$\gamma_2^{\rm ord}$ estimates are close but do not agree with the $S=100$ results. Taking instead
the most recent QMC estimate~\cite{sen_velocity_2015} for this ratio
$\rho_s/v=0.10882(4)$, the agreement is
clearly better. We have also checked that results on $\gamma_q^{\rm
  ord}(\ell/L)$ obtained by taking into account higher orders in $1/S$ from
Ref. \onlinecite{hamer_spin-wave_1994} give indeed an improvement over the $1/S$ order, but
are not as good as the QMC result.

\subsection{Connection to the free scalar field result}
\label{sec:gamma2}
\begin{figure}[b]
    \centering
    \includegraphics{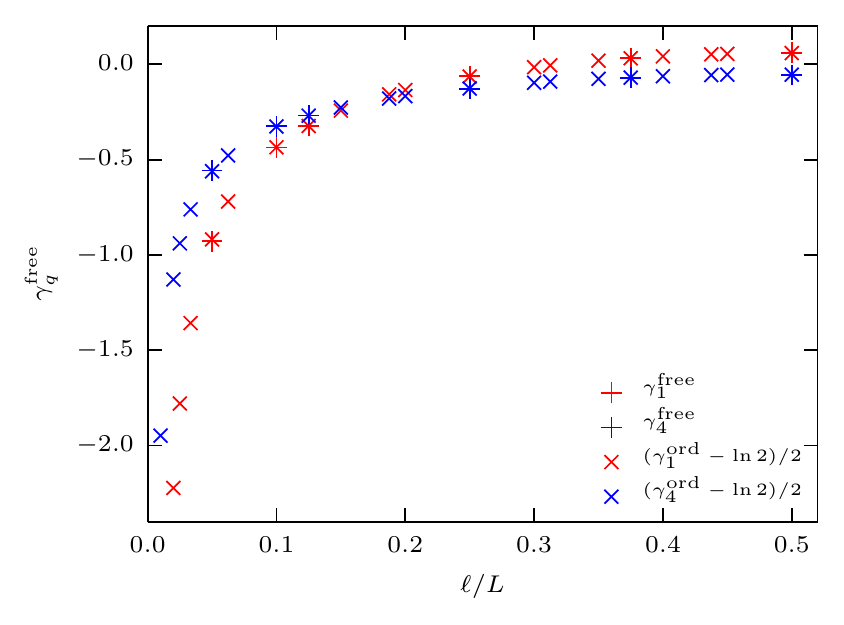}
    \caption{Universal additive constant $\gamma_q^{\rm free}(\ell/L)$ for free bosons plotted against the aspect ratio of the strip subsystem. Linear SW results for the $S=1/2$ Heisenberg antiferromagnet at $J_2=0$ are shown together with large-$S$ estimates of $\gamma_q^{\rm ord}$.}
    \label{fig:gammafree}
\end{figure}

In a (corner free) strip subsystem geometry with a finite aspect ratio $\ell/L$, one expects for gapped free bosons with a very large correlation length $\xi\gg L$ the following subleading corrections to the area law \cite{metlitski_entanglement_2011}:
\be
\Delta S_q=\frac{1}{2}\ln\left(\frac{\xi}{L}\right)+\gamma^{\rm free}_{q}(\ell/L),
\ee
where $\gamma^{\rm free}_{q}(\ell/L)$ is a universal geometric constant which depends non-trivialy on both the R\'enyi parameter and the aspect ratio.
By artificially gapping the linear SW Hamiltonian Eq.~\eqref{eq:quad} with a very small staggered field $h$, the dispersion relation in the vicinity of its two minima reads
\be
\Omega({\vec k})=\sqrt{8SJ_1 h+8S^2J_1(J_1-2J_2)|{\veck}|^2},
\ee
thus leading to 
\be
\xi=\sqrt{\frac{S(J_1-2J_2)}{h}}.
\ee
The correction due to the two minima becomes 
\be
\Delta S_q=\ln\left(\sqrt{\frac{S(J_1-2J_2)}{hL^2}}\right)+2\gamma^{\rm free}_{q}(\ell/L),
\label{eq:LS}
\ee
which is used to fit numerical SW results with a very small field $h=10^{-18}$ to extract $\gamma_{q}^{\rm free}(\ell/L)$, shown in Fig.~\ref{fig:gammafree}.

Quite interestingly, from the above formulation we can infer a very simple and direct relation between
$\gamma_{q}^{\rm ord}$ and $\gamma_{q}^{\rm free}$.  Indeed, in the large
$S$ limit the size-dependent staggered field (added to artificially restore
zero sublattice magnetization) takes the exact form
$h^*(L)=\frac{2J_1}{SL^4}$. Plugging this into Eq.~\eqref{eq:LS} yields
\be
\gamma_{q}^{\rm ord~SU(2)}=2\gamma_{q}^{\rm free} +\ln 2,
\ee
which agrees with MG~\cite{metlitski_entanglement_2011}, but only when $q=2$.
In Fig.~\ref{fig:gammafree}, comparing $\gamma_q^{\rm free}$ to $(\gamma_{q}^{\rm ord~SU(2)}-\ln 2)/2$ for $q=1, 4$ gives a perfect agreement for a wide range of aspect ratios.
 
One can also repeat the same argument for the XY model with only one Nambu-Goldstone mode~\cite{luitz_universal_2015} to get
\be
\gamma_{q}^{\rm ord~U(1)}=\gamma_{q}^{\rm free} +\frac{5}{4}\ln 2.
\ee
The reason for the disagreement between our large $S$ approach --- expected to become unbiased in the limit $S\to
\infty$ --- and the result from Ref.~\onlinecite{metlitski_entanglement_2011} for $\gamma_{q}^{\rm ord}$ is not clear at the moment.
\subsection{Limit of vanishing aspect ratio}
\label{sec:gamma3}
%

In this section we shed light on the divergent behavior of $\gamma^\text{ord}$ for small aspect
ratios by calculating $\gamma^\text{ord}$ in the extreme limit of $\ell/L \to 0$ using
subsystems with a fixed number $\ell$ of lines and thus a varying aspect ratio as a function of $L$. 
In order to achieve this, we work with $S=100$ at $J_2=0$, and we want to subtract all dominant terms, in particular the linear
area law contribution $a_q L$.

\begin{figure}[t]
    \centering
    \includegraphics{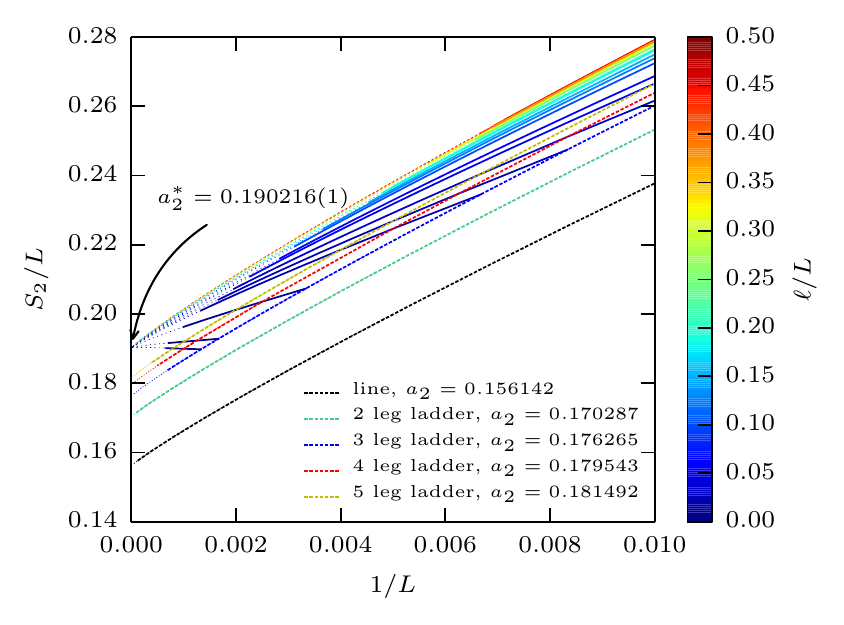}
    \caption{$S_2/L$ \vs $1/L$ for the strip shaped subsystem with different aspect ratios (rainbow
        curves). SW data obtained for $S=100$ and $J_2=0$.
        The dotted lines show an extrapolation for infinite system sizes which demonstrates that for all
aspect ratios the leading scaling term is indeed an area law. We also show results for the line and
$2, 3, 4$ and $5$ leg ladder subsystems (labeled lines) which clearly deviate significantly from this behavior. This
discrepancy is caused by the singular behavior of $\gamma^\text{ord}$ (and higher subleading terms), as the aspect ratio of
these subsystems changes constantly with $L$ and runs into the divergence of small aspect ratios
(see text).
}
    \label{fig:area_aspect}
\end{figure}

Let us therefore start with a study of the dominant scaling contribution of $S_q$ by plotting
$S_q/L$ \vs $1/L$ as displayed in Fig. \ref{fig:area_aspect}.
We show the area law behavior by
plotting $S_2/L$ \vs $1/L$ and an extrapolation $L \to \infty$, which guarantees to eliminate all
subleading terms. The figure shows two sets of curves. In the first one, each curve corresponds to
subsystems with a constant aspect ratio, such that $\gamma_2^\text{ord}$ is a constant for each
curve. These curves all yield identical area law prefactors $a_2^* =0.190216(1)$ as expected.

The second set of curves shows results corresponding to a fixed number $\ell$ of lines in the subsystem
(\ie a $\ell$-leg ladder), which implies that the aspect ratio of the
subsystem is a function of $L$. The dominant linear prefactor $a_q^\ell$ found for the
$\ell$-leg ladder subsystem is different from the fixed aspect ratio value, which is approached only for $\ell\gg 1$.
The reason for this discrepancy lies in the divergent behavior of
$\gamma_2^\text{ord}$ when $\ell/L$ tends to zero and the fact that the assumption that the only
surviving term in the scaling of $S_q/L$ at large sizes is the area law is no longer true. In fact, as the aspect
ratio of the subsystem changes constantly, $\gamma_q^\text{ord}$ seems to have a contribution that
is linear in the inverse aspect ratio and hence leads to a shift or an effectively \emph{changed area law
prefactor}. As a next step, we will try to determine this contribution.

\begin{figure}[t]
    \centering
    \includegraphics{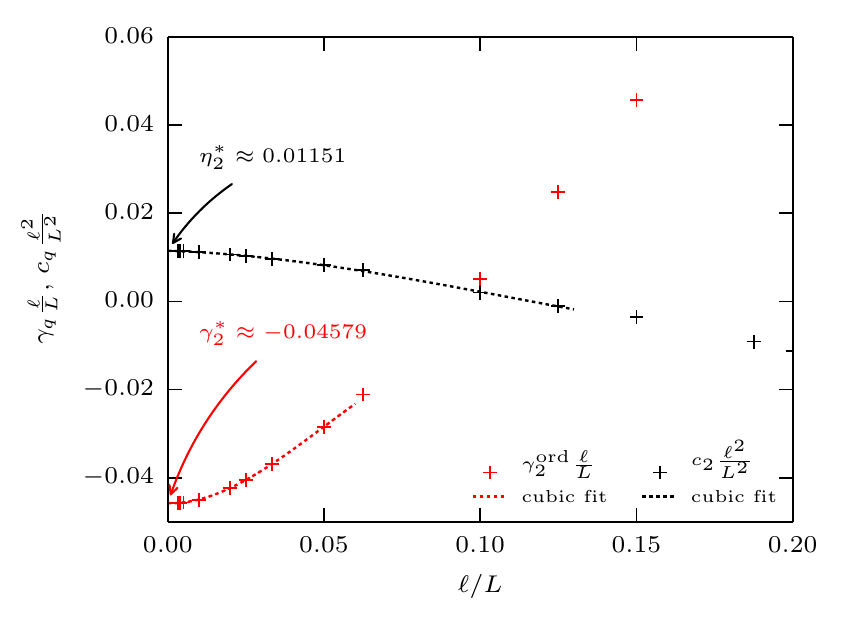}
    \caption{Singular contribution $\gamma_2^*$ of $\gamma_2^\text{ord}(\frac{\ell}{L})$ (red) and
    singular contribution $\eta_2^*$ of $c_q$ (black). The dashed 
line corresponds to cubic fits for the smallest aspect ratios and are used to extract estimates of
$\gamma_2^*$ and $\eta_2^*$, given by the intercept at vanishing aspect ratio. $\gamma_2^*$ corresponds to the
contribution of $\gamma_2$ to the linear scaling of the entanglement entropy of fixed width
subsystems. $\eta_2^*$ is the linear scaling contribution stemming from the EE scaling term
$c_q/L$.}
    \label{fig:gamma_singular}
\end{figure}

Fig \ref{fig:gamma_singular} shows our data for $\gamma_2^\text{ord}$ as obtained in Sec.
\ref{sec:gamma} multiplied by the aspect ratio $\frac{\ell}{L}$ as a function of the aspect ratio in
order to extract the singular contribution $\gamma^*_2$ as the intercept at vanishing aspect ratio.
The data shows convincing evidence that this contribution indeed extrapolates to a nonvanishing
value, which we determine by a cubic fit. With this information, we can now decompose
$\gamma_q^\text{ord}$ in a singular and a regular component:

\bea
    &\gamma_q^\text{ord}&\left(\frac{\ell}{L}\right) =  
    \gamma_q^\text{reg,ord}\left(\frac{\ell}{L}\right)+ \frac{L}{\ell} \gamma_q^*, \nonumber\\
    &\lim&_{\frac{\ell}{L}\to 0}\,\left(\frac{\ell}{L} 
    \gamma_q^\text{reg,ord}\right) = 0.
    \label{eq:gamma_singular_regular}
\eea
For completeness, we provide a table of $\gamma_q^*$ for other R\'enyi indices $q$ in Tab.
\ref{tab:gammastar}.

\begin{table}[b] \centering \begin{tabular}{c|cccc}
        \hline
        \hline
        $q$ & 1 & 2 & 3 & 4 \\
        \hline
        $\gamma_q^*$ & $-0.07677$ & $-0.04579$ &  $-0.03530$ & $-0.03123$  \\
        \hline
        \hline
    \end{tabular} 
    \caption{Values of $\gamma_q^*$ for different
    R\'enyi indices.} 
    \label{tab:gammastar} 
\end{table}

In general, we can assume that other subdominant terms show pathologic behavior in the limit of
vanishing (and non constant) aspect ratios, \ie for fixed width $\ell$ subsystems, we will for the
moment assume that they could produce a total correction to the area law of the form $\eta^*
\frac{L}{\ell^2}$ in total. The scaling of the EE then reads:

\bea
    S_q &=& \left( a^*_q + \gamma^*_q/\ell + \eta_q^*/\ell^2 \right) L \nonumber\\
    &+& \frac{n_G}{2} \ln \left(
    \frac{\rho_s}{v} L \right) + \gamma_q^\text{reg,ord} + \dots
    \label{eq:EEscaling_singular}
\eea
Clearly, for fixed aspect ratio subsystems the terms $\gamma_q^*$ and $\eta_q^*$ become irrelevant
for the area law for large system sizes. However, for fixed width subsystems, the effective linear
(in $L$) scaling coefficient $a_q^\ell$ is in fact given by

\begin{equation}
    a_q^\ell = a_q^* + \gamma_q^*/\ell + \eta_q^*/\ell^2.
    \label{eq:aq_ell}
\end{equation}
We can therefore obtain (in the limit of large $L$) $\gamma_q^\text{ord}$ from fixed width
subsystems by subtracting several terms from $S_q$: Obviously we need to subtract $\left( a_q^\ell  -
\gamma_q^*/\ell \right) L$ to eliminate the linear contribution (note how this automatically takes
care of the unknown terms $\eta_q^*$).

The second term that we have to subtract from the EE is the logarithmic term which is due to the
spontaneous breaking of SU(2) symmetry. We have argued above alongside with several works
\cite{metlitski_entanglement_2011,song_entanglement_2011,kulchytskyy_detecting_2015} that its value is $n_G/2=1$ for the case of fixed aspect ratio subsystems and
shown in Ref. \onlinecite{luitz_universal_2015} that this is also true for fixed width $\ell$
subsystems, such as the single line with $\ell=1$, we therefore subtract the term $n_G/2
\ln(\frac{\rho_s}{v} L)$, taking also care of the constant stemming from $\rho_s/v$, that we know
with great accuracy for the case of $S=100$ at $J_2=0$.

Remaining subleading terms are expected to die off in the limit of $\ell/L \to 0$ and are therefore
unimportant in the region of interest.

In total, for the limit of $L\to \infty$ and a fixed width $\ell$ of the subsystem, we obtain
$\gamma_q^\text{ord}$ through:

\begin{equation}
    \gamma_q^\text{ord}\left(\frac{\ell}{L}\right) = S_q - \left( a_q^\ell - \gamma_q^*/\ell \right)L -
    \frac{n_G}{2} \ln \left(\frac{\rho_s}{v}L\right).
    \label{eq:gamma_from_fixed_width}
\end{equation}

\begin{figure}[t]
    \centering
    \includegraphics{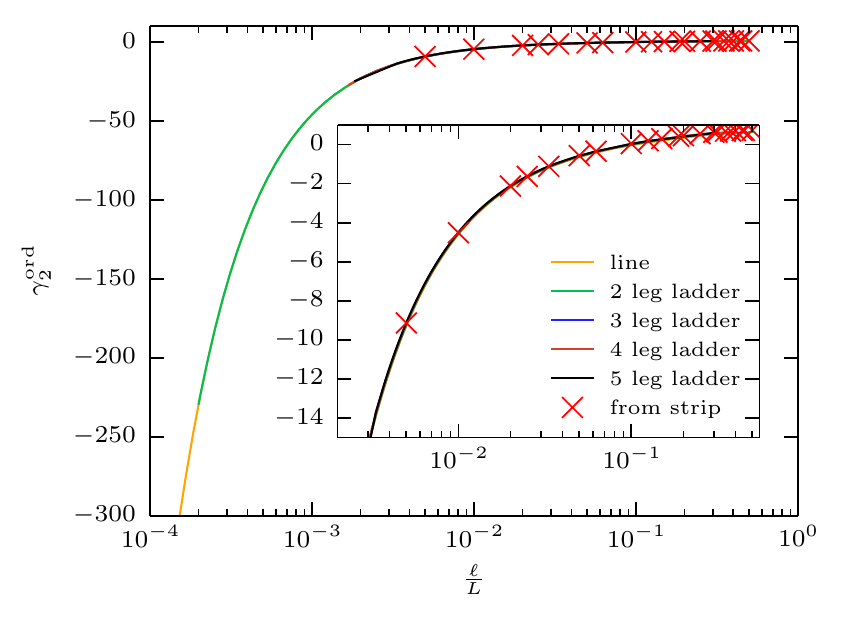}
    \caption{Data for $\gamma_2^\text{ord}$ as obtained from fits for fixed aspect ratio subsystems
    (red x) shown together with the result of Eq. \eqref{eq:gamma_from_fixed_width} for fixed width
subsystems. The agreement is excellent, even for relatively large aspect
ratios which
correspond to small systems in the fixed width case. The inset shows a zoom. Lines are guides to the
eye.}
    \label{fig:gamma_line_final}
\end{figure}

We can now apply Eq. \eqref{eq:gamma_from_fixed_width} to calculate $\gamma_q^\text{ord}$ in the
small aspect ratio regime from fixed width subsystems of width $\ell$. Fig.
\ref{fig:gamma_line_final} shows our result in comparison the previously obtained values of
$\gamma_2^\text{ord}$ from fixed aspect ratio subsystems (strips).  SW results for $\ell=1$ and $\ell=2$ are built on an analytical derivation (presented in Appendix~\ref{sec:1d}) obtained exploiting the fully symmetric nature of such subsystems. The perfect agreement of the
results obtained with different methods and in particular the agreement of the results for different
$\ell$ is a strong evidence for the reliability of this result and therefore demonstrates also the
singular nature of $\gamma_q^\text{ord}$ given by the singular component $\gamma_q^*$.

\begin{table}[hb]
    \centering
    \begin{tabular}{l|ccc}
        \hline
        \hline
        $\ell$ & $a_q^\ell$ & $\quad a_q^* + \gamma_q^*/\ell + \eta^*_q/\ell^2\quad$ &  $a_q^* + \gamma_q^*/\ell$ \\
        \hline
        1 & 0.156142 & 0.155936 & 0.144426  \\
        2 & 0.170287 & 0.170199 & 0.167321 \\
        3 & 0.176265 & 0.176232 & 0.174953  \\
        4 & 0.179543 & 0.179488 & 0.178768 \\
        5 & 0.181492 & 0.181518 & 0.181058 \\
        \hline
    \end{tabular}
    \caption{Comparison of the directly obtained linear scaling factor $a_q^\ell$ of fixed width
    subsystems to the result obtained using the singular contributions $\gamma_2^*$ and $\eta_2^*$
from subdominant terms. The inclusion of $\eta^*_q$ significantly improves the result and provides
numerical evidence for the correctness of Eq. \eqref{eq:aq_ell}. }
    \label{tab:aqell}
\end{table}

Can higher subleading terms generate corrections to the area law coefficient? It is certainly legal
to assume that pathological behavior in the limit of $\ell/L \to 0$ is not only present in the
scaling constant $\gamma_q^\text{ord}$ but also in higher terms, such as $c_q/L$ and $d_q/L^2$.
However, for them to modify the area law coefficient, they have to diverge much faster, \ie as
$\ell^2/L^2$ for the case of $c_q$. In order to investigate this possibility, we plot $c_2
\frac{\ell^2}{L^2}$ in black in Fig. \ref{fig:gamma_singular} and observe that a (small) nonzero
contribution to the linear scaling in $L$ is indeed present which we call $\eta_2^*$ (here we will
neglect the contributions to $\eta$ from even higher terms, which are difficult to access through
fits to numerical data). Let us finally plug all the information together and see if the singular
contributions of subdominant terms can explain the discrepancy between
$a_q^*$ and $a_q^\ell$ observed in Fig. \ref{fig:area_aspect} by
comparing in Tab. \ref{tab:aqell} results for $a_q^\ell$ as obtained from a direct fit to fixed
width EEs and for $a_q^* + \gamma_q^*/\ell + \eta^*_q/\ell^2$. The left
column shows the total linear scaling prefactor $a_q^\ell$ for fixed width subsystems as displayed
in Fig. \ref{fig:area_aspect}, while the rightmost column shows the fixed aspect ratio linear
scaling prefactor $a_q^*$ corrected by the singular contribution of $\gamma_q^\text{ord}$, giving
reasonable agreement. The middle column takes into account the next subdominant singular
contribution $\eta_q^*$ from the term $c_q/L$ as discussed above and reproduces the direct fit
result to very high accuracy, thus providing strong evidence for the correctness of Eq.
\eqref{eq:aq_ell}. We expect that even less dominant terms, such as $d_q/L^2$ will provide further
corrections, which are relevant for small widths $\ell$ and should account for the remaining
discrepancy, these terms are however very small and extremely difficult to extract numerically.

\section{Discussions and conclusions}
\label{sec:5}
In this work, we have performed a high-precision SW study of the $J_1 - J_2$ Heisenberg SU(2) antiferromagnet on the square lattice in order to investigate its quantum entanglement properties. Numerical calculations on finite size systems have been performed with an artificial restoration of zero sub-lattice magnetization using a small size-dependent staggered field $h^*(L)$. Several situations have been explored, and we have obtained finite size scaling results at large enough size such that the various terms appearing in the entanglement entropies have been precisely computed. 

The universal logarithmic correction due to Nambu-Goldstone modes associated with the breaking of a continuous symmetry (SU(2) in the present case) are well captured, giving a correction perfectly fitted by $\frac{n_G}{2}\ln L$, independent of the R\'enyi index $q$. In the case of subsystem having sharp corners, additional (negative) logarithmic corrections have been precisely evaluated, in perfect agreement with scalar field  theory predictions~\cite{casini_universal_2007}. The $J_1 -J_2$ model also offers a nice playground where we could check universality of the logarithmic term across the entire ordered regime but where we could further study the non-universal area law part which exhibits a non-trivial behavior, with a noticeable growth approaching the critical point in the frustrated regime. In the opposite limit of strong ferromagnetic second neighbor coupling, the mean-field limit is recovered with a vanishing area law term and a smooth crossover to a purely logarithmic scaling of the entropies.

Part of this work was also devoted to the study of the additional constant term $\gamma_{q}^{\rm ord}$, expected to be universal for strip subsystems~\cite{metlitski_entanglement_2011}, only depending on their aspect ratio. It then appeared crucial to impose zero sublattice magnetization in the finite size SW theory, yielding a unique size-dependent staggered field $h^*(L)$ which (i) mimics a tower of state gap $\sim 1/L^2$ in the excitation spectrum (responsible for the logarithmic correction), and (ii) leads to the correct additional geometric constant $\gamma_{q}^{\rm ord}$, in perfect agreement with MG~\cite{metlitski_entanglement_2011}, at least for $q=2$. A simple and direct relation with non-interacting bosons was also derived. Finally we have precisely investigated the limit of vanishing aspect ratios using very large ladder subsystems in the limit of finite number of legs, discovering that the geometric constant contains both a regular part and a singular component in this limit.
Our study is concluded by showing that singular components of even less dominant terms explain
perfectly the discrepancy of the area law terms obtained from fixed width \vs fixed aspect ratio
subsystems.

Among the potentially interesting future directions, a quantitative study of the geometric constant
using exact Monte Carlo, while very challenging, appears to be a very important point in order to
test the validity of our prediction for $q>2$. It may also be interesting to extend the present SW
approach to other continuous symmetries like SU(N) models using modified flavor-wave theory for instance. Other geometries or $d=3$ are certainly of great interest also, with a larger choice of subsystem shapes.

\acknowledgments{We would like to thank T. Grover, R. Melko, M. Metlitski, and T. Roscilde for discussions. We are grateful to H. Casini for sharing with us the estimates of the corner contributions from Ref.~\onlinecite{casini_universal_2007}. We also want to acknowledge X. Plat for participation
  in related projects. This work was performed using HPC resources from GENCI (Grant No. x2015050225) and CALMIP (Grant No. 2015-P0677),
and is supported by the French ANR program ANR-11- IS04-005-01.}
\bibliography{EE}

\newpage

\appendix
\section{Details of spin-wave calculations}
\label{sec:appA}

This appendix provides details of spin-wave calculations which are not
crucial for the computation of entanglement entropy, but which are
nonetheless useful for an understanding of the method and its range of
validity. We also provide a comparison between the finite-size SW approach
and direct QMC computations for $S=1/2$ for the antiferromagnetic structure factor in the
ferromagnetic range of next neighbor coupling $J_2<0$.

\subsection{Spin-wave spectrum and velocity}

\begin{figure}[h] \centering \includegraphics[width=.9\columnwidth,clip]{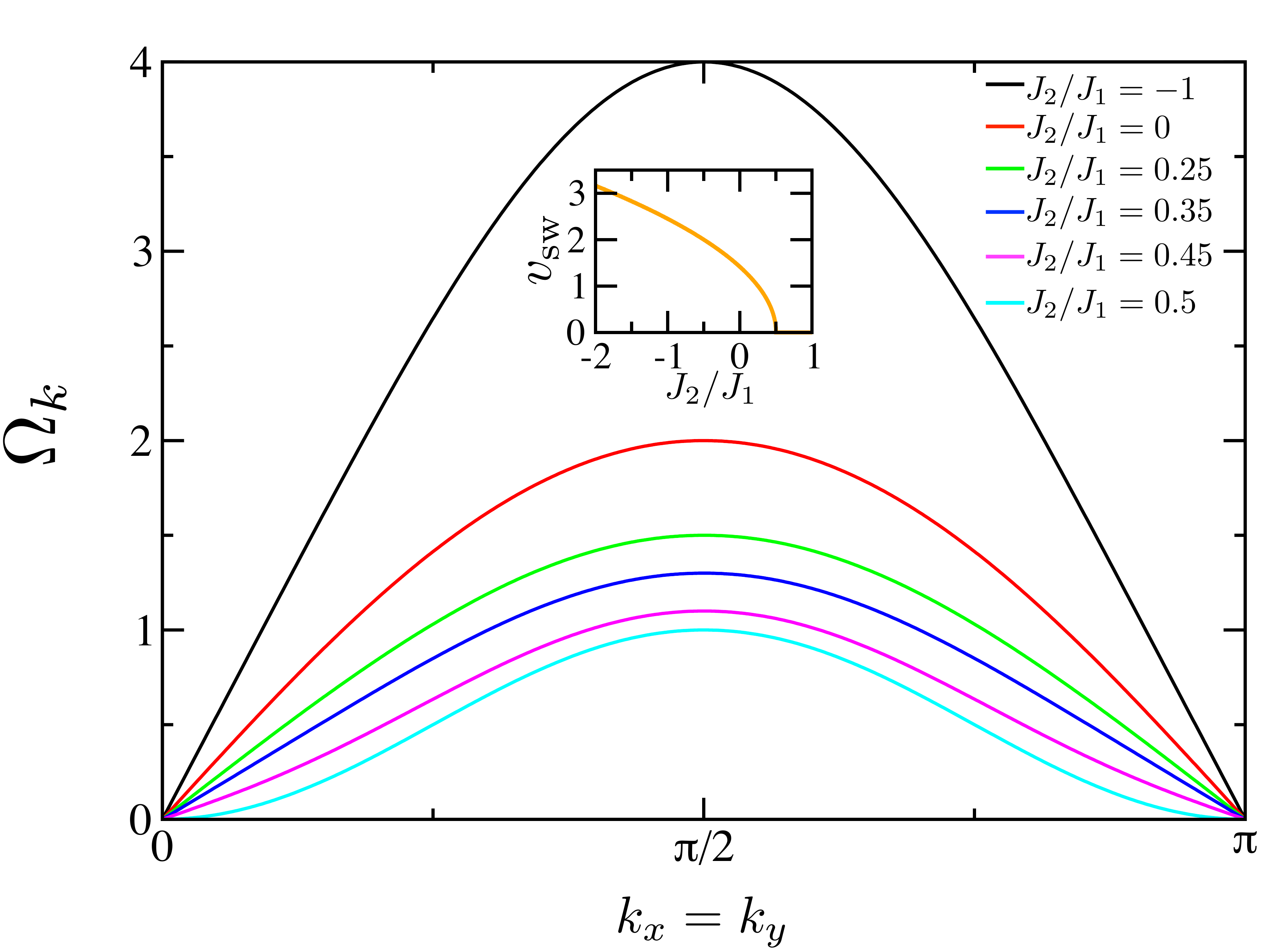}
\caption{SW spectrum at $h=0$ for various $J_2/J_1$ plotted along the $k_x=k_y$ direction. Inset: SW     velocity $v_{\rm sw}$.}
\label{fig:SW} \end{figure}

\begin{figure}[h] \centering \includegraphics[width=.8\columnwidth]{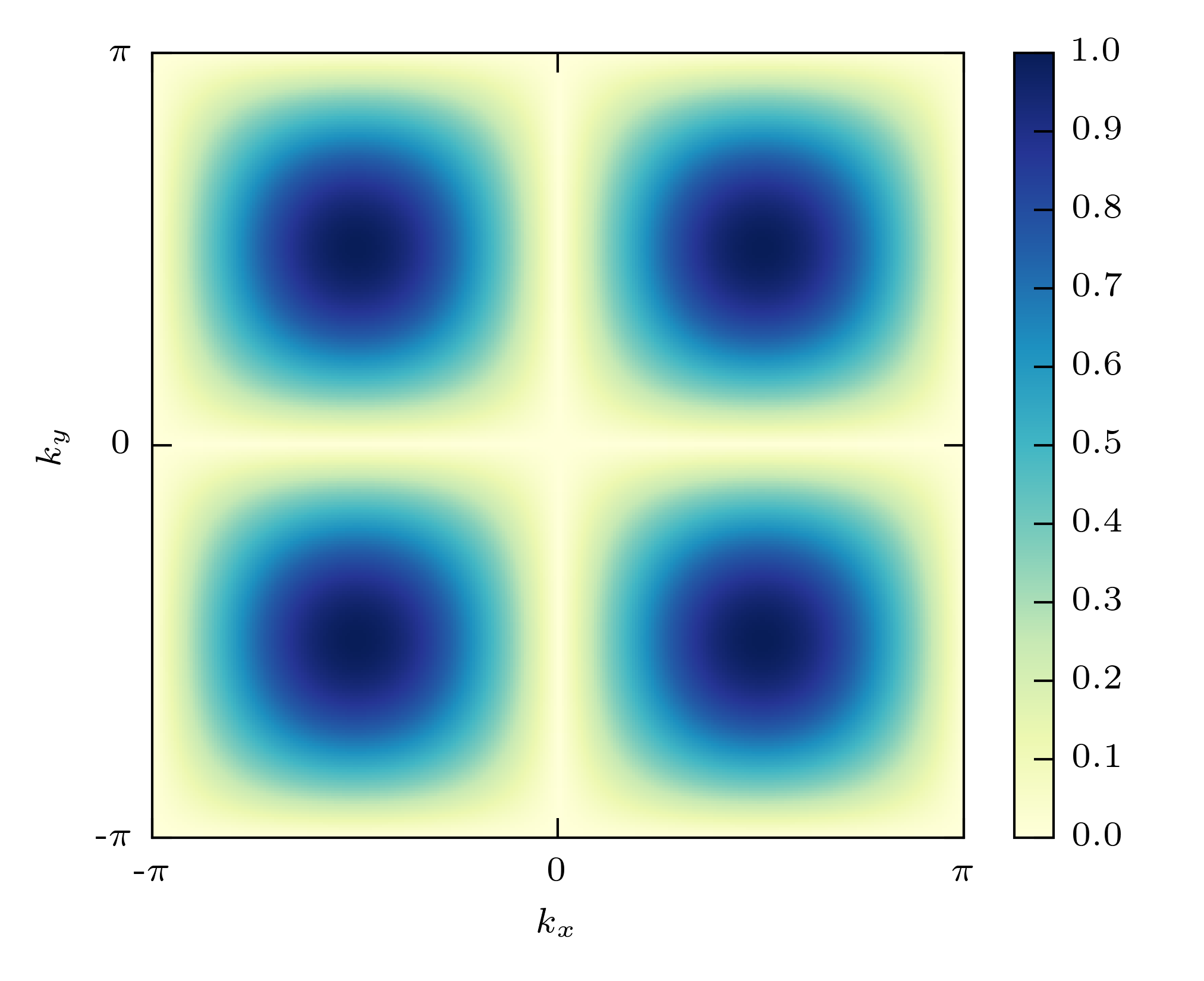}
\caption{2D color map of the SW spectrum at $h=0$ for $J_2/J_1=1/2$ and $S=1$. A line of minima is visible
    along the $k_x=0$ and $k_y=0$ directions.  }
\label{fig:map} \end{figure}

We present in Fig.~\ref{fig:SW} the spectrum $\Omega_{\bf k}=2\sqrt{A_{\bf k}^2 -B_{\bf k}^2}$ in the
direction $k_x=k_y$ (obtained from expressions Eq.~\eqref{eq:AkBk}) as a function of $k_x$, for
different coupling strenghts $J_2$ and for a spin value $S=1/2$. The
inset represents the spin-wave velocity, Eq.~\eqref{eq:vsw}, as a function of
$J_2/J_1$. We see that the velocity vanishes at $J_2/J_1=0.5$ where the SW spectrum features a continuous line of minima at $k_x=0$ and $k_y=0$, as depicted in Fig.~\ref{fig:map}.

\subsection{Range of non-vanishing staggered magnetization}
\subsubsection{Antiferromagnetic order parameter}

\begin{figure}[h] \centering \includegraphics[width=\columnwidth,clip]{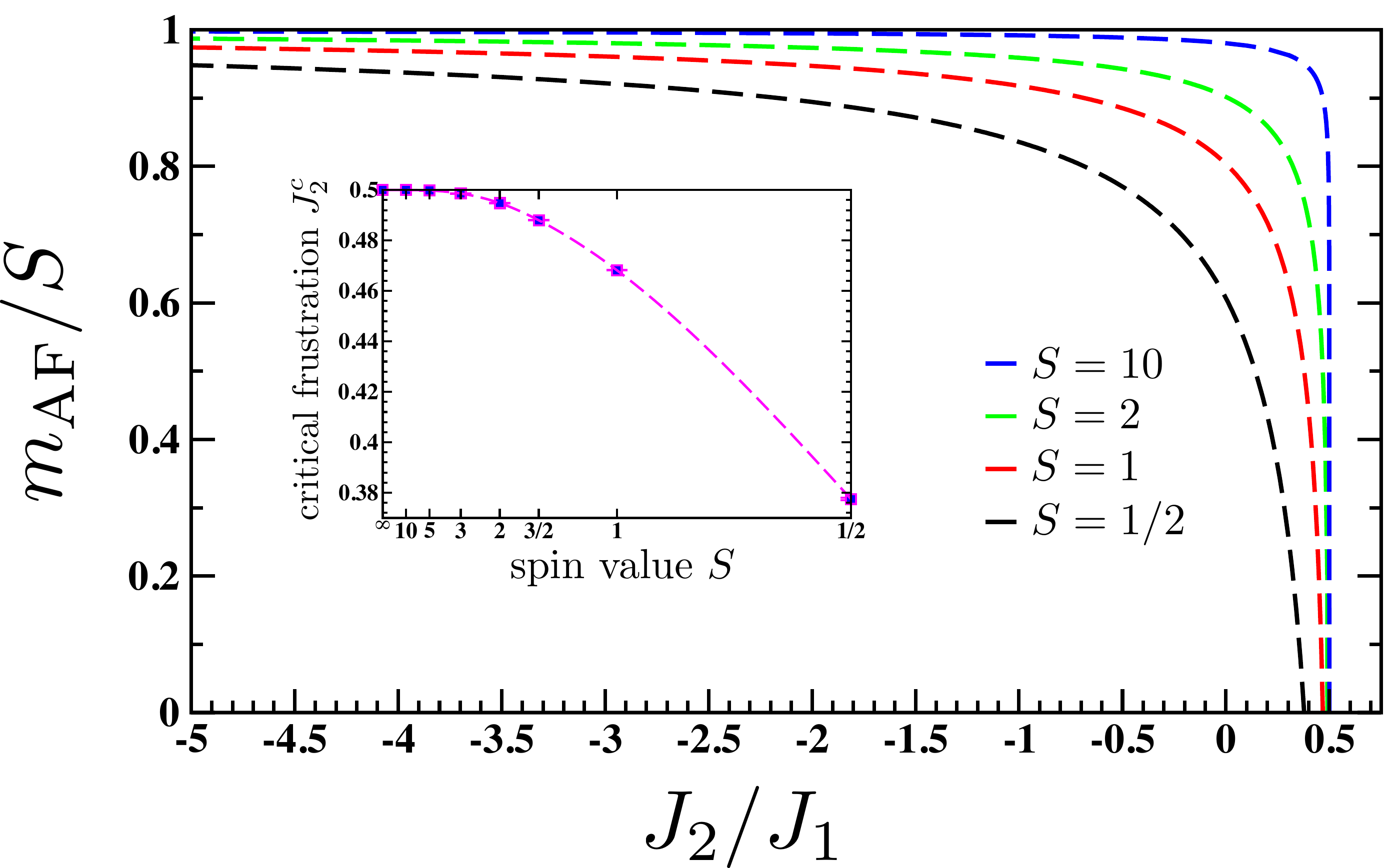}
\caption{SW results for the AF order parameter Eq.~\eqref{eq:mafsw} of the $J_1 - J_2$ model on the square lattice for various spin sizes $S$. Inset: Critical frustration $J_2^c$ (in units of $J_1$) plotted against the spin length $S$.}
\label{fig:mafsw} \end{figure}

Eq.~\eqref{eq:mafsw} can be evaluated numerically for different values of the
spin size $S$ and second neighbor coupling strength $J_2/J_1$, in order to probe the
range of validity of the spin-wave approach, which assumes an ordered
ground-state. This is illustrated in Fig.~\ref{fig:mafsw} where the AF order
parameter is represented, and as expected, is clearly
enhanced by ferromagnetic diagonal coupling $J_2/J_1<0$ while it decreases
towards zero when $J_2/J_1$ approaches $1/2$. The critical frustration
$J_2^c$ (in units of $J_1$), above which the SW-corrected order parameter
vanishes, is also represented in the inset of Fig.~\ref{fig:mafsw} as a
function of $S$ where we observe that $J_2^c\to 1/2$ when $S$ gets larger.

\subsubsection{Finite size SW: AF structure factor}
\begin{figure}[t] \centering \includegraphics[width=\columnwidth,clip]{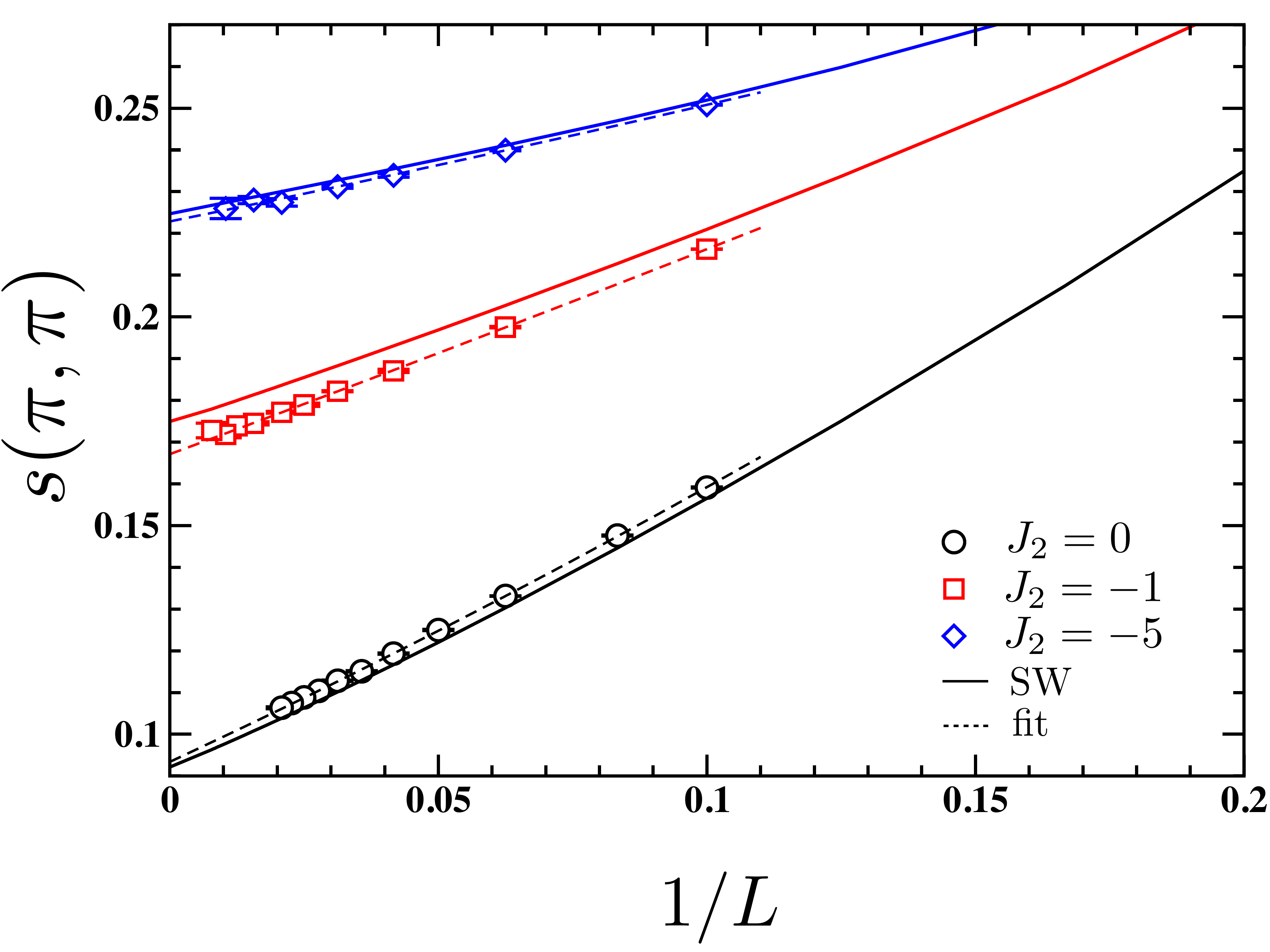}
	\caption{Staggered structure factor per site $s(\pi,\pi)$ of the $J_1 - J_2$ antiferromagnet Eq.~\eqref{eq:J1J2} plotted against the inverse system length $1/L$ for 3 values of $J_2$, with $J_1=1$. Symbols show $T=0$ QMC results, dashed lines are quadratic fits of the form Eq.~\eqref{eq:spipiL}, and the full lines are modified SW results using Eq~\eqref{eq:spipisw}.}
\label{fig:spipi} \end{figure}

To illustrate the interest of using a formulation of SW which treats more
correctly finite-size systems, we present results for the computation of the
staggered structure factor per site on finite square lattices $L\times L = N$:
\be
s(\pi,\pi)=\frac{1}{N^2}\sum_{ij}(-1)^{|i-j|}\langle {\bf S}_i\cdot{\bf S}_j\rangle.
\label{eq:spipi}
\ee
Using Wick's theorem, all two-spin correlators can be computed in terms of
the $f_{ij}$ and $g_{ij}$ functions defined in Eq.~\eqref{eq:fijgij} of the
main text. Imposing that $\langle {\bf S}_i\cdot{\bf S}_j\rangle = \langle
{S}_i^z{S}_j^z\rangle $ (because the theory is strictly speaking not rotationally invariant), we obtain:
\be
{s}(\pi,\pi)=\frac{1}{N^2}\sum_{ij}\left(f_{ij}^2+g_{ij}^2\right)-\frac{1}{4N}.
\label{eq:spipisw}
\ee
A quantitative comparison between the above SW expectation and exact quantum Monte Carlo simulations is shown in Fig.~\ref{fig:spipi}. Ground-state expectation values for $s(\pi,\pi)$ of the $J_1 - J_2$ Hamiltonian Eq.~\eqref{eq:J1J2} with $S=1/2$ and $J_2=0,~-1,~-5$ have been obtained for various square lattices $L\times L$ using the stochastic series expansion algorithm~\cite{syljuasen_quantum_2002}.
One sees in Fig.~\ref{fig:spipi} that the agreement is fairly good, in particular for strong second neighbor ferromagnetic coupling $J_2/J_1=-5$. Interestingly, the finite size scaling behavior, expected from previous works~\cite{huse_ground-state_1988,sandvik_finite-size_1997} 
\be
s(\pi,\pi)=m_{\rm AF}^2+m_1/L +m_2/L^2+\cdots
\label{eq:spipiL}
\ee
is very well captured by SW calculations, as visible in Table~\ref{tab:1} where QMC and SW estimates for $m_{\rm AF},~m_1$, and $m_2$ are compared and show a good agreement.
\begin{table}[h!]
\begin{tabular}{|c|l|l|l|}
\hline
$J_2/J_1$&$m_{\rm AF}^2$&$m_1$&$m_2$\\
&SW~/~QMC&SW~/~QMC&SW~/~QMC\\
\hline
$~~0$&$0.092~/~0.093(1)$&$0.55~/~0.60(1)$&$0.8~/~0.6(1)$\\
$-1$&$0.175~/~0.167(1)$&$0.42~/~0.47(2)$&$0.4~/~0.15(9)$\\
$-5$&$0.225~/~0.223(1)$&$0.25~/~0.26(2)$&$0.2~/~0.2(1)$\\
\hline
\hline
\end{tabular}
\caption{\label{tab:1}Fit parameters from Eq.~\eqref{eq:spipiL}.}
\end{table}
%
%

The fact that finite size corrections are well captured by this modified SW
formalism is a confirmation that it is a good starting point to study
ground-state properties on finite systems and in particular the finite size
scaling of the entanglement entropy, as discussed in the main text.

\section{Mean-field limit}
\label{sec:MF}
    \begin{figure}[b] \centering \includegraphics[width=\columnwidth,clip]{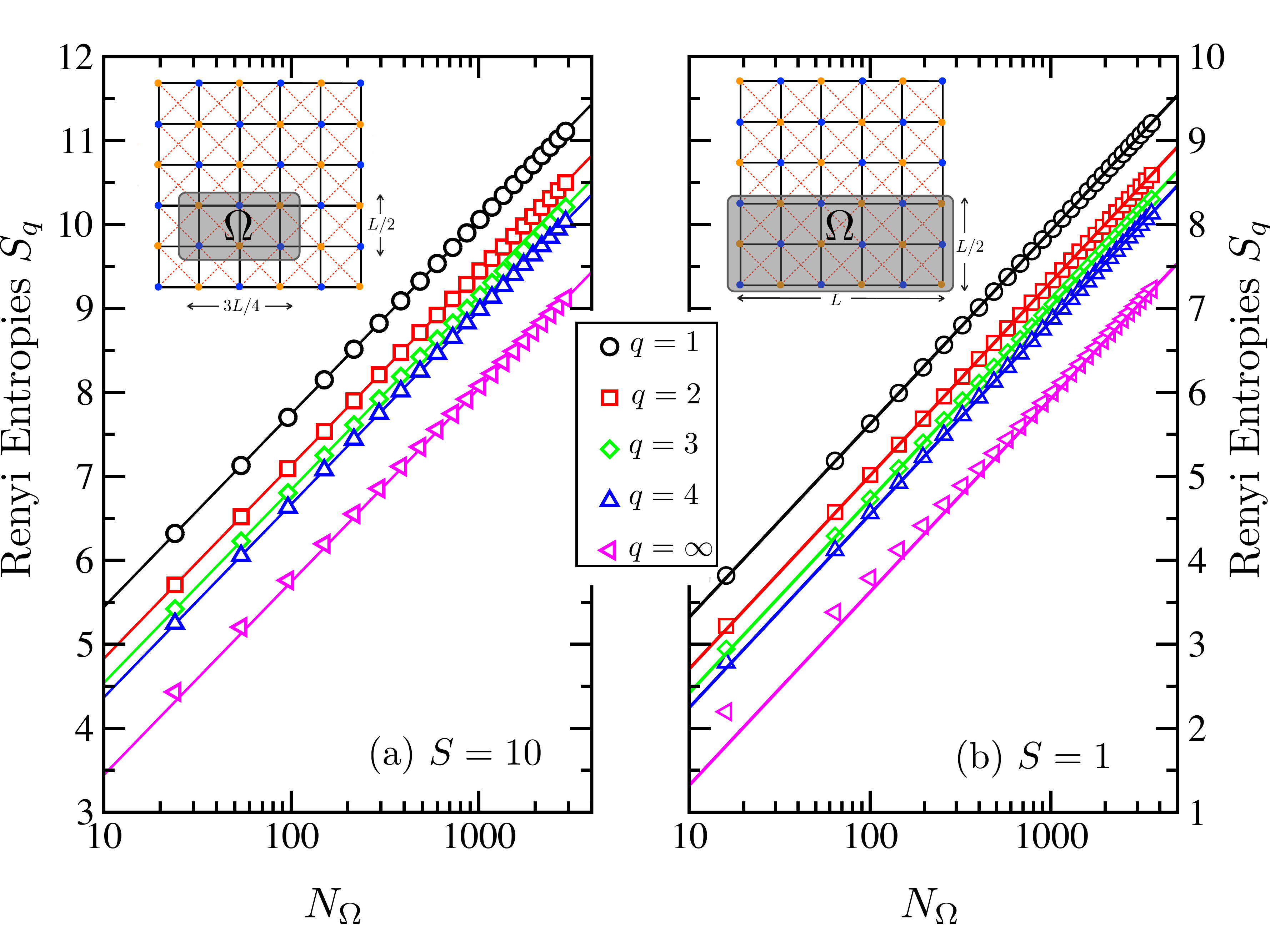}    
    	\caption{Mean-field limit for the R\'enyi entropies $S_q$. The symbols show numerical results for $J_2/J_1=-10^5$ with different geometries and spin lengths ($S=1,~10$), plotted against the number $N_{\Omega}$ of sites in the subsystem $\Omega$. The numerical results (symbols) are compared to the analytical expression Eq.~\eqref{eq:MF} with  $r_{\Omega}=3/8$ (a) or $r_\Omega=1/2$ (b), shown by the full lines.}
\label{fig:MF} 
\end{figure}
In the limit $-J_2/J_1\gg 1$ one should recover the mean-field result obtained for example for the Lieb-Mattis model~\cite{vidal_entanglement_2007,ding_block_2008}. 
In such a limit, perfect ferromagnetic correlations between spins belonging to the same sublattice imply $f_{ij}=S$ for $i\neq j$ both on the same sublattice ($g_{ij}=0$) and $f_{ii}=S+1/2$. Antiferromagnetic correlations between opposite sublattices yield
\be
\sum_{ij}f_{ij}-g_{ij}=0,
\ee
thus leading to $g_{ij}=S+1/N$ for $i\neq j$ on opposite sublattices ($f_{ij}=0$).
Therefore non-zero matrix elements of the correlation ${\bf C}$ (for $i$ and $j$ on the same sublattice) are given by
\bea
C_{ii}&=&S(1-r_{\Omega})+\frac{1}{4}\nonumber\\
C_{ij}&=&S(1-r_{\Omega}),
\eea
where $r_{\Omega}=N_{\Omega}/N$ is the ratio between the number of sites
inside the sub-system $N_{\Omega}$ and the total number of sites $N$. The
spectrum of the correlation matrix ${\bf C}$ is then straigthforwardly given by
\bea
\nu^2_{1,2}&=&\frac{N_{\Omega}}{2}S(1-r_{\Omega})+\frac{1}{4}\\
\nu^2_{l}&=&\frac{1}{4},~~~~~(l=3,\ldots,N_{\Omega}).
\eea
one sees that only two eigenvalues contribute, in a  macroscopic way.
We then compute directly the R\'enyi entropies for any partition $N_{\Omega}$ and any $q\ge 1$
\be
S_q=\ln (N_{\Omega}) +\ln\left[\frac{S(1-r_{\Omega})}{2}\right]+2\frac{\ln q}{q-1}.
\label{eq:MF}
\ee
The area law term vanishes, and the dominant scaling is now a pure
logarithm of the number of sites $N_{\Omega}$. This exact expression can be
compared to the numerical solution of the SW Hamiltonian for very large negative values of $J_2$. In Fig.~\ref{fig:MF} we show numerical results for $J_2/J_1=-100000$ for two values of $S$ and two different geometries, which compares extremely well with the MF limit expression Eq.~\eqref{eq:MF}. Note that the lines are not fitting functions. If we try instead to fit to the general form $a_q L + l_q \ln L + b_q + c_q/L +d_q/ L^2$, we end up with $a_q\sim 10^{-11}$ and $l_q=2$.

\begin{figure}[h]
    \centering
    \includegraphics{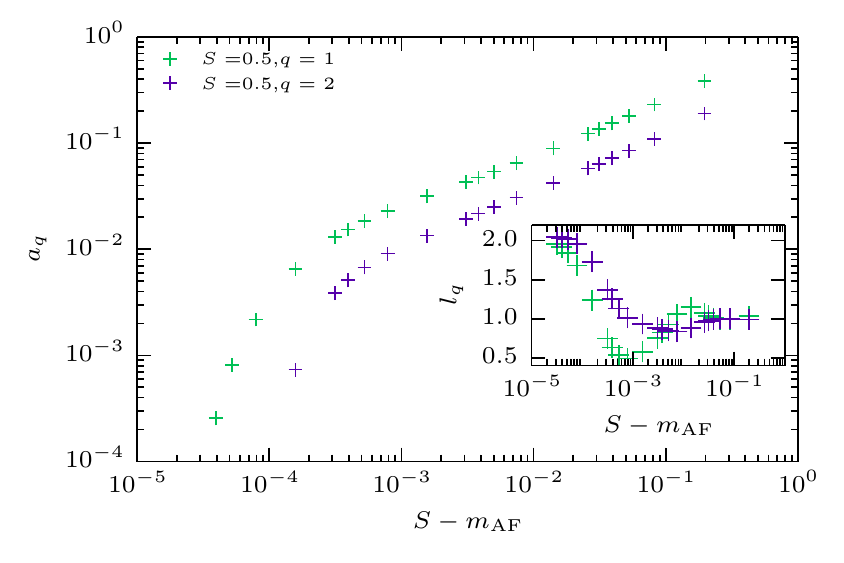}
    \caption{Area law coefficient $a_q$ as a function of the normalized order parameter
        $S-m_\text{AF}$, which is small in the ordered phase. Here, we show results for
        $S=\frac{1}{2}$. The inset depicts the prefactor of the logarithmic entanglement entropy
    scaling term, which has a plateau at $l_q=1$ for intermediate $J_2$ and evolves to the mean
field limit of $J_2\rightarrow -\infty$. It is unclear if the behavior in the crossover region is a
finite size effect or an artefact of the spin wave method.}
    \label{fig:arealaw_maf}
\end{figure}

Before concluding, we want to briefly comment on the crossover to the MF limit when the ferromagnetic second neighbour is turned on towards very large values. This is illustrated in Fig.~\ref{fig:arealaw_maf} where the rapid decrease of the area law coefficient $a_q$ (for $q=1,2$) is shown versus the quantum depletion of the AF order parameter $S-m_{\rm AF}$. In the same time, the log coefficients $l_1$ and $l_2$, plotted in the inset of Fig.~\ref{fig:arealaw_maf}, crossover from $l_q=1$ up to $l_q=2$ in the limit of vanishing quantum fluctuations $m_{\rm AF}\to S$. 
\section{Analytical derivation for one-dimensional subsystems}
\label{sec:1d}
A great simplification for the computation of entanglement entropy is possible if all sites $i$ and $j$ inside a
subsystem $\Omega$ are equivalent, or in other words if the matrix elements
$C_{ij}$ only depend on the relative distance
$\left|{\bf{r}}_i-{\bf{r}}_j\right|$. In such a case, for sites on different
sublattices $C_{ij}=0$. This situation is achieved for one-dimensional
subsystems with one or two lines (Fig.~\ref{fig:lattice} (a) with $\ell=1,2$). 
In these specific situations, we can derive analytic expressions for the eigenvalues of ${\bf{C}}$, avoiding a numerical diagonalization. This has first been discussed in Ref.~\onlinecite{luitz_universal_2015}, and we provide here details of this calculation, starting with the case of a line-shaped subsystem.

This subsystem being invariant under translations along the
$x$ direction, the functions $f_{ij}$ and $g_{ij}$ defined in
Eqs.\eqref{eq:fijgij} only depend on the distance $x=x_i-x_j$ along the subsystem. They reduce
consequently to
\bea
f_x & = & \frac{1}{2N} \sum_{k_x} \cos(k_x x) \alpha_{k_x}, \nonumber \\
g_x & = & \frac{1}{2N} \sum_{k_x} \cos(k_x x) \beta_{k_x} 
\eea
with 
\be
\alpha_{k_x}  :=  \sum_{k_y} \frac{A_\vec{k}}{\Omega_{\vec{k}}} \quad {\rm
  and} \quad \beta_{k_x}  := \sum_{k_y} \frac{B_\vec{k}}{\Omega_{\vec{k}}} 
\label{eq:alphakx}
\ee
which satisfy the property $\alpha_{k_x+\pi}=\alpha_{k_x}$, $\beta_{k_x}  = -\beta_{k_x+\pi}$.
Since the functions $f_x$ and $g_x$ possess translation and reflection symmetries 
\be
f_x=f_{L-x}=f_{L+x}, \quad \text{and} \quad g_x = g_{L-x}= g_{L+x},
\ee
so does the correlation matrix: $C_{ij}=C(l=|x_i-x_j|)=C(L-l)=C(l-L)$. Since
furthermore $C(x)$ vanishes for odd distances, it is convenient to re-index all sites on one sublattice from
$1$ to $L/2$ (say, blue sites in Fig.~\ref{fig:lattice}a for $\ell=1$), and
sites on the other sublattices from $L/2+1$ to $L$ (say, orange sites in Fig.~\ref{fig:lattice} a) to block-diagonalize $\mat{C}$ onto two identical blocks of size $L/2 \times L/2$. The translation invariance ensures that each block is {\it circulant}, with matrix elements $C(l)= \sum_{x\,\text{even}} f_x f_{x-l} - \sum_{x\,\text{odd}}
g_x g_{x-l}$. The eigenvalues $\nu_l^2$ of $\mat{C}$ are given by the properties of circulant matrices
\cite{gray_toeplitz_2005}:
\be
\begin{split}
    &\nu_l^2 = c(0) + (-1)^l c\left(\frac{L}{2}\right) + \sum_{j=1}^{\lceil L/4-1 \rceil} c(2j) \cos\left( \frac{4\pi}{L} j l
\right),\\
&l\in\{0,1,\dots \frac{L}{2}-1\},\, \, c\left(\frac{L}{2}\right)=0 \,\,\, \text{if} \,\,\,
\frac{L}{2}\,\text{mod}\,2=1.
\end{split}
\ee

We can even simplify calculations by noticing that $f_x$ and $g_x$ are
discrete Fourier transforms of $\alpha_k$ and $\beta_k$ respectively. Using the convolution theorem on $C(l)$,  we arrive at the final expression  for the $L$ eigenvalues of $\mat{C}$ of the {\it single-line} subsystem:
\be
\nu_q^2 = \frac{1}{4 N} (\alpha_q^2 - \beta_q^2),
\ee
with $q\in\{-\pi+\frac{2\pi}{L},\dots \pi\}$.

A very similar reasoning can be applied to the case of a 2-line (ladder)
subsystem with $2L$ sites ($\ell=2$ in Fig.~\ref{fig:lattice}a). It is convenient to re-index sites by labelling (in
a zig-zag fashion) all (say, blue in in Fig.~\ref{fig:lattice}a) sites of one
sublattice from $1$ to $L$, and (orange in Fig.~\ref{fig:lattice}a) sites
from the other sublattice from $L+1$ to $2L$. Again, $\mat{C}$ is block-diagonal with
identical circulants blocks with matrix elements $C(l)=\sum_{x=0}^{L-1}
{f}_x^+ {f}_{x-l}^+ - \sum_{x} \tilde{g}_x^+ \tilde{g}_{x-l}^+$ with
${f}_x^+=f(x,0)+f(x,1)$ and ${g}_x^+=g(x,0)+g(x,1)$. We can now introduce
the discrete Fourier tranforms of the newcomers $f(x,1)$ and $g(x,1)$
\be
\tilde{\alpha}_k:=\sum_{k_y} \frac{A_{\vec{k}} \cos (k_y)}{\Omega_{\vec{k}}} \quad {\rm and} \quad \tilde{\beta}_k:=\sum_{k_y} \frac{B_{\vec{k}} \cos (k_y)}{\Omega_{\vec{k}}}
\ee
 to again be able to apply the convolution theorem. We finally obtain that the $L$ eigenvalues of {\it one block} of $\mat{C}$ for the {\it ladder} subsystem are given by:
\be
\nu_q^2 = \frac{1}{4 N} \left[(\alpha_q-\tilde{\alpha}_q)^2 - (\beta_q-\tilde{\beta}_q)^2\right],
\ee
with $q\in\{-\pi+\frac{2\pi}{L},\dots \pi\}$. Since $\mat{C}$ has two identical blocks, the $2L$ eigenvalues for the ladder subsystem are obtained by doubling the above spectrum.

\end{document}